\documentclass{emulateapj}
\shorttitle{JEANS ANALYSIS AND DARK MATTER}
\shortauthors{S{\'A}NCHEZ-SALCEDO ET AL.}

\begin{document}
\title{Jeans analysis of the Galactic thick disk and the local dark matter density }
\author{F.~J.~S\'anchez-Salcedo\altaffilmark{1}, Chris Flynn\altaffilmark{2} 
and J.~A.~de Diego\altaffilmark{1}}
\altaffiltext{1}{Instituto de Astronom\'ia, Universidad Nacional Aut\'onoma
de M\'exico, Ciudad Universitaria, 04510 Mexico City, Mexico;
jsanchez@astro.unam.mx}
\altaffiltext{2}{Centre for Astrophysics and Supercomputing,
Swinburne University of Technology,
Australia}

\begin{abstract}
Dynamical estimates of the mass surface density at the solar radius can
be made up to a height of $4$ kpc using thick disk stars as tracers of the potential. 
We investigate why different Jeans estimators of the local
surface density lead to puzzling and conflicting results.
Using the Jeans equations, we compute the vertical ($F_{z}$) 
and radial ($F_{R}$) components of the gravitational force, as well as $\Gamma(z)$,
defined as $\Gamma\equiv\partial V_{c}^{2}/\partial R$, with $V_{c}^{2}\equiv -RF_{R}$.
If we assume that the thick disk does not flare and that 
all the components of the velocity dispersion tensor of the thick disk
have a uniform radial scalelength of $3.5$ kpc, $\Gamma$ takes 
implausibly large negative values, when using the currently available 
kinematical data of the thick disk. This implies that the input parameters
or the model assumptions must be revised. 
We have explored, using a simulated thick disk, the impact of the
assumption that the scale lengths of the density and velocity dispersions
do not depend on the vertical height $z$ above the midplane.
In the lack of any information about how these scale radii depend on $z$,
we define a different strategy.
By using a parameterized Galactic potential, we find that 
acceptable fits to $F_{z}$, $F_{R}$ and $\Gamma$ are obtained for a 
flaring thick disk and a spherical dark matter halo with a local density 
$\gtrsim 0.0064 M_{\odot}$pc$^{-3}$. Disk-like dark matter
distributions might be also compatible with the current data of the thick disk.
A precise measurement of $\Gamma$ at the midplane could be very useful
to discriminate between models.

\end{abstract}

\section{Introduction}
It is now generally accepted that our Galaxy contains a dark matter (DM)
halo with a virial mass between $0.6\times 10^{12}M_{\odot}$ and 
$3\times 10^{12}M_{\odot}$ (e.g., Wang et al. 2015 and references therein).
A wide range of experiments are currently ongoing or are planned aiming
to detect DM particles by direct scattering
between DM  and nuclei in detectors or indirectly
by their emission of secondary particles from DM annihilations 
(e.g., Bernabei et al.~2010; Angloher et al.~2012;
Aalseth et al.~2013; Agnese et al.~2013; Aartsen et al.~2015). 
Since the probability of collisions between DM particles and
detectors depends on the flux of DM particles, i.e.~on the
phase-space density, it is crucial to infer
the distribution function of DM particles in the Solar system
to estimate the chances of direct detection.
Based on the disk rotation curve and assuming that DM particles are
distributed in a quasi-spherical halo, one infers a typical DM
density of $\sim 0.01 M_{\odot}$pc$^{-3}$ at the
solar position, and a (one-dimensional) velocity dispersion of 
$\sim 150$ km s$^{-1}$.
However, the density structure of the halo could be more complex: 
Adiabatic response of the dark halo
to the baryonic component or the capture of satellite halos in low-inclination
orbits could lead to the formation of a thick dark disk superimposed on
the quasi-spherical halo (Read et al.~2008, 2009; Pillepich et al.~2014; 
Ruchti et al.~2014; Piffl et al.~2015).
A dark disk may enhance direct detection because the flux
of particules is proportional to the density. If the dark disk
is in counter-rotation, the flux is enhanced due to a larger
relative velocity between the Sun and the DM particles.

In recent years, many attempts have been carried out to determine the local
DM density within a few hundred parsecs of the Sun (see Read 2014 for a review).
Several authors have combined data from a wide range of tracers,
including the H\,{\sc i} rotation curve, and 
determined the local DM density $\rho_{0}$ by fitting
a global model for the Milky Way (Catena \& Ullio 2010; Weber \& de Boer 2010;
Iocco et al.~2011; McMillan 2011; Piffl et al.~2014). 
All these studies are consistent
with $\rho_{0}$ between $0.005$ and $0.015 M_{\odot}$pc$^{-3}$ within
$1\sigma$. Other groups have derived $\rho_{0}$, independently
of the rotation curve, calculating the gravitational potential 
up to a height of $1-1.5$ kpc, from an equilibrium distribution of tracer 
stars in the solar neighborhood.
From the kinematics of K stars, Garbari et al.~(2012) derived a
density of $0.022\pm^{0.014}_{0.013}M_{\odot}$pc$^{-3}$.
Zhang et al.~(2013), using K dwarfs, measured
$\rho_{0}=0.0065\pm 0.0023M_{\odot}$pc$^{-3}$.
Values of $\rho_{0}=0.008\pm 0.0025M_{\odot}$pc$^{-3}$ were derived 
in Bovy \& Rix (2013) by modeling the dynamics of G-type dwarfs.
Bienaym\'e et al.~(2014) using red clump stars up to a height of $2$ kpc,
derived $\rho_{0}=0.0143\pm 0.0011 M_{\odot}$ pc$^{-3}$. 
In most cases, the results for $\rho_{0}$ overlap within their stated
uncertainties. 
Only the $1\sigma$ interval of the recent measurement of 
Bienaym\'e et al.~(2014) does not overlap; 
if the quoted error bars are not underestimated, this 
may suggest the existence of potential sources of systematics.

The contribution of the dark halo should clearly manifest at large
heights from the midplane. However,
Moni Bidin et al.~(2010, 2012b) 
carried out a Jeans analysis of a sample
of thick disk stars up to a height of $4$ kpc and found no need for
DM to account for the observations 
($\rho_{0}=0\pm 0.001M_{\odot}$pc$^{-3}$).
Bovy \& Tremaine (2012) reanalysed their data using other
model assumptions, finding values fully consistent with standard 
estimates of this quantity.
The estimate of $\rho_{0}$ depends on the adopted radial and vertical
scalelengths of the thick disk, denoted by $h_{R}$ and $h_{z}$, 
respectively. In particular, for $h_{R}=2$ kpc and $h_{z}=0.7$ kpc, 
Bovy \& Tremaine (2012) obtained $\rho_{0}\geq 0.0095\pm 0.0015M_{\odot}$pc$^{-3}$,
whereas for $h_{R}=3.8$ kpc and $h_{z}=0.9$ kpc, they found
$\rho_{0}\geq 0.007\pm 0.001M_{\odot}$pc$^{-3}$.
More recently, Moni Bidin et al.~(2015) have reconsidered their 
three-dimensional formalism and have found that 
$\rho_{0}=0\pm 0.002 M_{\odot}$pc$^{-3}$ for $h_{R}=2$ kpc and $h_{z}=0.7$
kpc and   $\rho_{0}=0.002\pm 0.003 M_{\odot}$pc$^{-3}$ for
$h_{R}=3.6$ kpc and $h_{z}=0.9$ kpc. 
It is remarkable that, even using
the same data, the estimates on the local DM density
in Bovy \& Tremaine (2012) and Moni Bidin et al.~(2015)
differ by one order of magnitude for $h_{R}=2$ kpc and 
$h_{z}=0.7$ kpc.

In this paper, we examine why the analyses of
Bovy \& Tremaine (2012) and Moni Bidin et al.~(2015) 
can give different results and explore 
the impact of uncertainties in each approach.  Then, we suggest a 
more robust alternative to estimate the surface density,
which takes into account information of the kinematics in the 
radial direction.  

The paper is organized as follows.
In \S \ref{sec:basic}, we outline the basis and assumptions behind the
Jeans analysis and present the one-dimensional approach used by
Bovy \& Tremaine (2012) and the three-dimensional formalism advocated
by Moni Bidin et al.~(2015). In \S \ref{sec:MW}, we compare the
predictions of these two estimators of the surface density when 
using the kinematics of thick disk stars. 
In order to gain physical insight on the different terms and their
uncertainties, we compute the vertical and radial components of 
the gravitational force, as well as the radial term in the Poisson
equation, using the Jeans equations, and compare with the values
inferred in representative mass models. 
In \S \ref{sec:sims}, we use some mock data from numerical simulations
to identify any possible bias in the assumptions. 
In \S \ref{sec:parametricandmodel}, we calculate the local
parameters of the dark halo,
using a parametric method, under the assumption that the DM halo
is spherically-symmetric within the solar circle.
Conclusions are given in \S \ref{sec:conclusions}.

\section{Surface density estimators}
\label{sec:basic}
\subsection{Jeans analysis and  Poisson equation}
\label{sec:Jeans_Poisson}
The kinematics of tracer stars can be used to determine the 
gravitational potential of an astronomical object.
The Jeans equations provide the strength of the components of
the gravitational force from the kinematics of tracer stars. 
For an axisymmetric steady-state disk with mean velocities 
$\bar{V}_{R}=\bar{V}_{z}=0$ in cylindrical coordinates, the
vertical and radial components of the gravitational force are given by:
\begin{equation}
F_{z}=\frac{1}{\nu}\left[\frac{\partial (\nu\sigma_{z}^{2})}{\partial z}
+\frac{\nu \sigma_{Rz}^{2}}{R}+\frac{\partial (\nu\sigma_{Rz}^{2})}{\partial R}\right],
\label{eq:fz}
\end{equation}
and
\begin{equation}
F_{R}=\frac{1}{\nu}\frac{\partial(\nu\sigma_{R}^{2})}{\partial R} +\frac{1}{\nu}
\frac{\partial (\nu\sigma_{Rz}^{2})}{\partial z}+\frac{\sigma_{R}^{2}-
\sigma_{\phi}^{2}-\bar{V}_{\phi}^{2}}{R},
\label{eq:fR}
\end{equation}
where $\nu(R,z)$ is the volume mass density of the tracer population,
$\sigma_{ij}^{2}=\overline{V_{i}V_{j}}-\bar{V}_{i}\bar{V}_{j}$ is its
velocity dispersion tensor and
$\bar{V}_{\phi}$ is its mean velocity in the azimuthal direction.
To simplify notation, we use $\sigma_{i}^{2}\equiv \sigma_{ii}^{2}$.

Once $F_{z}$ and $F_{R}$ are known, we may estimate the total density 
of mass (baryonic plus DM) $\rho$, using the Poisson equation that relates 
the gradients of the gravitational force to $\rho$:
\begin{equation}
4\pi G\rho (R,z)=-\frac{\partial F_{z}}{\partial z}-\frac{1}{R}\frac{\partial}{\partial R}
(RF_{R}).
\end{equation}
Integration of the above equation between $-Z_{1}$ and $Z_{1}$
leads us to infer the total column density to $Z_{1}$  at a given
distance $R$: 
\begin{equation}
2\pi G\Sigma(R,Z_{1})=S_{F_{z}}+S_{F_{R}},
\label{eq:init}
\end{equation}
where
\begin{equation}
S_{F_{z}}\equiv -F_{z}(R,Z_{1}),
\end{equation}
and
\begin{equation}
S_{F_{R}}\equiv \frac{1}{R}\int_{0}^{Z_{1}}\Gamma(R,z) dz,
\label{eq:sfR}
\end{equation}
where 
\begin{equation}
\Gamma(R,z)\equiv \frac{\partial V_{c}^{2}}{\partial R},
\end{equation}
and $V_{c}^{2}(R,z)\equiv -RF_{R}$.
We will refer to $V_{c}$ as the ``circular velocity''.

We note that whereas $S_{F_{z}}$ is positive 
for a centrally condensed distribution of mass,  
$S_{F_{R}}$ may in general be either positive or negative.
If the mass distribution is very flattened and oblate, such as in a 
massive disk, $S_{F_{z}}$ is larger than $|S_{F_{R}}|$ at small enough $z$. 
For the potential created by a point-mass particle, we have 
$|S_{F_{R}}|=S_{F_{z}}$.
Finally, we may have $|S_{F_{R}}|\gtrsim S_{F_{z}}$ for a mass
distribution elongated along the $z$ axis (i.e.~prolate distribution).

$F_{z}$, $F_{R}$ and $\Gamma$, and thereby $S_{F_{z}}$ and $S_{F_{R}}$,
can be written in terms of the density $\nu$, the velocity dispersions
$\sigma_{ij}^{2}$, and their first and second derivatives
(see Appendix \ref{sec:second_order} for more details). 
While $S_{F_{z}}$ is essentially
the vertical force at the height of interest ($Z_{1}$), the
computation of the term $S_{F_{R}}$ requires knowledge of
the $R$-gradient of $V_{c}^{2}$ from $z=0$ to $z=Z_{1}$. Thus,
$S_{F_{R}}$ is in general more uncertain than $S_{F_{z}}$ because the 
computation of $S_{F_{R}}$ involves second order derivatives along 
the radial direction\footnote{Particularly 
uncertain is the radial derivative of $\bar{V}_{\phi}$
as a function of $z$, which is necessary to compute $S_{F_{R}}$.}.
Therefore, using the Poisson equation is the natural
path for deriving $\Sigma$ when having very small errors in the 
measured quantities $\nu(R,z)$,
$\sigma_{ij}^{2}(R,z)$ and $\bar{V}_{\phi}(R,z)$ of the tracer population. 
At low vertical heights and at distances where the rotation velocity curve 
is nearly flat, it holds that $|S_{F_{R}}|\ll |S_{F_{z}}|$ and 
thus uncertainties on $S_{F_{R}}$ will have a minimal
effect on the surface density estimate.

We consider the stars in the thick disk as our tracer population and
assume that its density distribution can be described by 
\begin{equation}
\nu \propto \exp \left(-\frac{R-R_{\odot}}{h_{R}}-\frac{|z|}{h_{z}}\right), 
\end{equation}
where $h_{R}$ and $h_{z}$ are the radial and vertical scalelengths, 
respectively (Siegel et al.~2002; Juri\'c et al.~2008; Bovy \& Rix 2013). 
The simplest case is to assume that the thick disk has uniform (constant)
scalelengths. However, 
in order to include a possible flare of the tracer disk 
(e.g., Mateu et al.~2011; Polido et al.~2013;
L\'opez-Corredoira \& Molg\'o 2014; Minchev et al.~2015), 
$h_{z}$ may depend on $R$; $h_{z}=h_{z}(R)$ and denote 
$\xi\equiv dh_{z}/dR$. On the other hand, it is likely that
the radial scalelength $h_{R}$ is not strictly constant with $z$. 
Here, we consider the generic case that $h_{R}(z)$.

Regarding the components of the velocity dispersion,
$\sigma_{ij}^{2}$, we assume that they all 
exponentially decay along $R$ with scalelength $h_{\sigma_{ij}}$ 
(Lewis \& Freeman 1989; Bovy et al.~2012a; Hattori \& Gilmore 2015):
\begin{equation}
\sigma^{2}_{ij}(R,z)=\sigma^{2}_{ij}(R_{\odot},z) \exp\left(-\frac{R-R_{\odot}}{h_{\sigma_{ij}}(z)}\right).
\end{equation}
We will refer to $h_{R}$, $h_{z}$ and $h_{\sigma_{ij}}$ as 
the ``geometrical'' parameters.
Note that the radial scalelengths of the velocity dispersion
tensor ($h_{\sigma_{ij}}$) may vary with $z$.

Under these approximations, the dynamical estimates of 
$F_{z}$ and $F_{R}$ at $R=R_{\odot}$, which are denoted
by $F_{z}^{\rm est}$ and $F_{R}^{\rm est}$, are 
\begin{equation}
F_{z}^{\rm est}=\frac{\partial \sigma_{z}^{2}}{\partial z}-\frac{\sigma_{z}^{2}}{h_{z}}
+k_{0}\sigma_{Rz}^{2},
\label{eq:fzest}
\end{equation}
with $k_{0}\equiv R^{-1}-h_{R}^{-1}-h_{\sigma_{Rz}}^{-1}+\xi z h_{z}^{-2}$ and
\begin{equation}
F_{R}^{\rm est}=k_{0}'\sigma_{R}^{2}-\frac{1}{R}(\sigma_{\phi}^{2}+\bar{V}_{\phi}^{2})
+\frac{\partial \sigma_{Rz}^{2}}{\partial z}-\frac{\sigma_{Rz}^{2}}{h_{z}},
\label{eq:fRest}
\end{equation}
with $k_{0}'\equiv R^{-1}-h_{R}^{-1}-h_{\sigma_{R}}^{-1}+\xi z h_{z}^{-2}$.

Under the same assumptions, the radial derivative of $V_{c}^{2}$ at 
$R=R_{\odot}$, which is required to compute $S_{F_{R}}$, is
\begin{equation}
-\Gamma^{\rm est}= k_{1}\sigma_{R}^{2}+
\frac{\sigma_{\phi}^{2}}{h_{\sigma_{\phi}}}+k_{2}R 
\left(\pm\frac{\sigma_{Rz}^{2}}{h_{z}}-\frac{\partial \sigma_{Rz}^{2}}{\partial z}\right)+\frac{\xi R}{h_{z}^{2}}\sigma_{Rz}^{2}
-\frac{\partial{\bar{V}_{\phi}^{2}}}{\partial R}
\label{eq:dVc_dR}
\end{equation}
where 
\begin{equation}
k_{1}=\left(1-\frac{R}{h_{\sigma_{R}}}\right)k_{0}'-\frac{1}{R}-
\frac{zR}{h_{z}^{2}}\left(\frac{2\xi^{2}}{h_{z}}-\frac{d\xi}{dR}\right),
\label{eq:k1k1}
\end{equation}
and $k_{2}=h_{\sigma_{Rz}}^{-1}-R^{-1}$.
The plus-minus sign within the parentheses in Equation (\ref{eq:dVc_dR}) 
indicates that the plus sign must be taken when $z>0$, and the
minus sign when $z<0$.

We stress that Equations (\ref{eq:fzest})-(\ref{eq:k1k1}) 
are valid at $R=R_{\odot}$. We have omitted extra terms of the form
\begin{equation}
(R-R_{\odot})\frac{\sigma_{ij}^{2}}{h_{\sigma_{ij}}^{2}} 
\frac{\partial h_{\sigma_{ij}}}{\partial z},
\end{equation}
because we are only interested in the vertical profiles of $F_{z}$,
$F_{R}$ and $\Gamma$ at the cylindrical galactocentric radius of the
Sun. At $R\neq R_{\odot}$, these terms should be taken into account.

\subsection{Bovy \& Tremaine's estimator}
\label{sec:btestimator}
Bovy \& Tremaine (2012) explored the magnitude of $S_{F_{R}}$
for three mass distributions: a single exponential disk with
a scalelength of $3.4$ kpc, a single NFW halo, and a combination of the
two in which the circular speed is flat at $R_{\odot}$. They found
that $0\leq S_{F_{R}}\leq 0.2 S_{F_{z}}$ within $|z|\leq 4$ kpc and
suggested, from Equation (\ref{eq:init}), that the formula
\begin{equation}
\Sigma_{BT}(Z_{1})=-\frac{F_{z}^{\rm est}(Z_{1})}{2\pi G},
\end{equation}
with $F_{z}^{\rm est}$ given in Equation (\ref{eq:fzest}) with $\xi=0$,
leads to  an underestimate of the surface density to $4$ kpc
only by $\sim 20\%$. We will refer to this formula
as the Bovy \& Tremaine (BT) estimator.  Note that, unlike the 
classical one-dimensional approximation (e.g., Read 2014),
the cross term of the velocity dispersion, $k_{0}\sigma_{Rz}^{2}$,
is included in the computation of $F_{z}^{\rm est}$. 

Moni Bidin et al.~(2015) warned that it may be misleading to assume
that $S_{F_{R}}(z)$ is positive for any Galactic mass model, since
it depends on the relative weight of the different mass components
of the Milky Way. If so, the BT estimator does not necessarily
yields a lower limit to the surface density, nor is it accurate
within $20\%$.  According to Moni Bidin et al.~(2015),
the assumption $S_{F_{R}}(z)>0$ is not adequate to derive $\rho_{0}$
because it is implicitly constraining the mass distribution.

\subsection{Moni Bidin et al.'s estimator}
To compute $\Sigma (Z_{1})$, Moni Bidin et al.~(2015) prefer to retain 
the term $S_{F_{R}}$ in Equation (\ref{eq:init}), and assume that all the
geometrical parameters of the thick disk are constant (i.e.~the disk
does not flare and the scalelengths $h_{R}$ and $h_{\sigma_{ij}}$ are 
independent of $z$). 
In such a case and combining Equations (\ref{eq:sfR}) and (\ref{eq:dVc_dR}), we obtain:
\begin{equation}
\Sigma(Z_{1})=\frac{1}{2\pi G}\left( -F_{z}^{\rm est}(Z_{1})+S_{F_{R}}(Z_{1})\right),
\label{eq:gen}
\end{equation}
where
\begin{eqnarray}
\label{eqn:mb14_correct}
&&S_{F_{R}}(Z_{1})=-\frac{k_{1}}{R}\int_{0}^{Z_{1}} \sigma_{R}^{2} dz- 
\frac{1}{Rh_{\sigma_{\phi}}}\int_{0}^{Z_{1}}
\sigma_{\phi}^{2}dz\\ \nonumber
&&+k_{2}  \left(\sigma_{Rz}^{2}-\frac{1}{h_{z}}\int_{0}^{Z_{1}} \sigma_{Rz}^{2}dz\right)
+\frac{2}{R}\int_{0}^{Z_{1}} \bar{V}_{\phi} 
\frac{\partial \bar{V}_{\phi}}{\partial R} dz.
\end{eqnarray}

All the terms in Eq.~(\ref{eqn:mb14_correct}) 
coincide with those in the equation used in 
Moni Bidin et al.~(2015) --their equation (11)-- except the term:
\begin{equation}
-\frac{k_{2}}{h_{z}}\int_{0}^{Z_{1}}\sigma_{Rz}^{2} dz,
\label{eq:missingterm}
\end{equation}
which does not appear in the equation for $\Sigma$
used by Moni Bidin et al.~(2015).
This term will be included 
in the present study.
In fact, we will show in \S \ref{sec:bt_vs_cMB} that this term may be 
important for some choices of the geometrical parameters.
We will refer to Equation (\ref{eq:gen}) together with
Equation (\ref{eqn:mb14_correct}) as the corrected Moni Bidin et al.~(cMB) 
estimator, which it will be denoted by $\Sigma_{cMB}$.

\section{Application to the Milky Way}
\label{sec:MW}
\subsection{The tracer population: Data and fits}
\label{sec:data}
In this Section, we compile the data used to study the different
estimators of the local surface density in the Milky Way.
As the tracer population, we use the Galactic thick disk.
Measurements of the velocity dispersions of $412$ thick
disk stars were provided by Moni Bidin et al.~(2012a). The observed velocity
dispersion components of stars, at $|z|>1.5$ kpc, in the thick disk are:
\begin{equation}
\sigma_{R}=(82.9\pm 3.2)+(6.3\pm 1.1) (|z|-2.5)\,\, {\rm km \,\,s^{-1}},
\label{eq:sigmaR}
\end{equation}
\begin{equation}
\sigma_{\phi}=(62.2\pm 3.1)+(4.1\pm 1.0) (|z|-2.5)\,\, {\rm km \,\,s^{-1}},
\label{eq:sigmaphi}
\end{equation}
\begin{equation}
\sigma_{z}=(40.6\pm 0.8)+(2.7\pm 0.3) (|z|-2.5)\,\, {\rm km \,\,s^{-1}},
\label{eq:sigmaz}
\end{equation}
(Moni Bidin et al.~2012a). 
The vertical profile of the cross term $\sigma_{Rz}^{2}$ is
very irregular at $z<4.5$ kpc. Moni Bidin et al.~(2010) fit it
by a linear function at $z>3$ kpc for which they obtained 
\begin{equation}
\sigma_{Rz,1}^{2}=(1522\pm 100)+(366\pm 30)(z-2.5)\,\, {\rm km^{2} \,\,s^{-2}}.
\label{eq:sigmaRz1}
\end{equation}
This fit was used by Moni Bidin et al.~(2012b, 2015) to make a 
dynamical inference of $\Sigma(z)$.
Since all the substructure in $\sigma_{Rz}^{2}$ is well resolved,
we include all the data points. Fitting
all the available data, we find, for $z>1.5$ kpc,
\begin{equation}
\sigma_{Rz,2}^{2}=(0\pm 100)+(450\pm 60)z\,\, {\rm km^{2} \,\,s^{-2}}.
\label{eq:sigmaRz2}
\end{equation}

To apply the cMB estimator,
we also need $\bar{V}_{\phi}$ and $\partial \bar{V}_{\phi}/\partial R$. 
We use 
\begin{equation}
\bar{V}_{\phi}=V_{c,0}-(22.5\pm 3)-(22.5\pm 3)|z|^{1.23\pm 0.03},
\label{eq:Vphi}
\end{equation}
where
$V_{c,0}$ is the rotational velocity in the midplane at $R_{\odot}$,
$z$ is given in kpc and $\bar{V}_{\phi}$ in km s$^{-1}$ 
(Moni Bidin et al.~2012a). The adopted values for $V_{c,0}$
will be specified later on. Finally,
for $\partial \bar{V}_{\phi}/\partial R$,
we have fitted the data collected by Moni Bidin et al.~(2015)
from the Sloan Digital Sky Survey (SDSS, York et al.~2000) and
the data of Casetti-Dinescu et al.~(2011), through a linear function.
We obtained 
\begin{equation}
\frac{\partial \bar{V}_{\phi}}{\partial R}=(4\pm 1)|z|+(1.4\pm 1.4).
\label{eq:dVphi_dR}
\end{equation} 
Current measurements of $\partial \bar{V}_{\phi}/\partial R$ are 
limited to $z\leq 2.7$ kpc. Strictly speaking, the inferences 
using the above fit for $\partial \bar{V}_{\phi}/\partial R$ are only 
valid at $z<2.7$ kpc. 

In their analysis to derive the velocity dispersions of the thick
disk stars (Eqs.~\ref{eq:sigmaR}-\ref{eq:sigmaRz2}), 
Moni Bidin et al.~(2012a) did not account for Poisson noise,
which is important due to the small size of the sample. For this reason,
Sanders (2012) pointed out that Moni Bidin et al.~(2012a) underestimated
the gradients of the velocity dispersions with Galactic height.
Moni Bidin et al.~(2015) argue that the gradient estimates quoted in Moni
Bidin et al.~(2012a) should be accurate within $15\%$ and that,
even enhancing the vertical gradients by a factor of three, the impact
on the resuls is small.
In order to have better inferences of the gradients,
better data are clearly necessary (Moni Bidin et al.~2015).

\begin{figure*} %
\epsscale{1.150}
\plotone{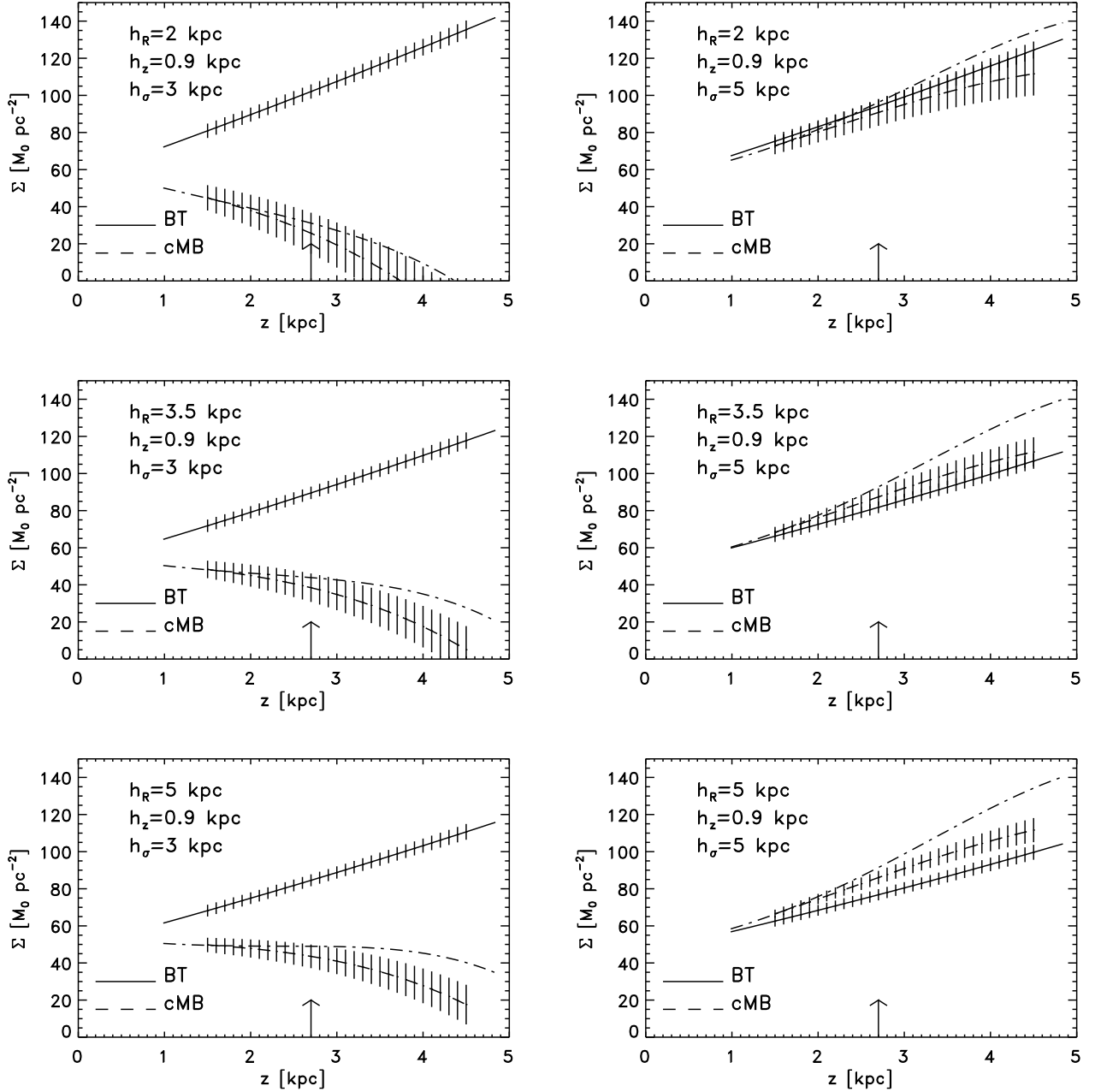}
\caption{Surface density $\Sigma(z)$ predicted in the Milky Way, at $R=R_{\odot}$,
using BT (solid lines) and cMB (dashed lines) estimators for 
different combinations of the geometrical parameters.  
In order to evaluate $\Sigma_{cMB}$, the observed values of
$\partial\bar{V}_{\phi}/\partial R$ were fit with a linear
(dashed lines) and with an $|z|^{1.5}$-function 
(dashed-dotted lines). At $|z|>2.7$ kpc, the values  
for $\partial\bar{V}_{\phi}/\partial R$ were extrapolated. 
The arrows mark this height.
  }
\label{fig:Moni_MW}
\end{figure*}

\subsection{BT estimator vs cMB estimator}
\label{sec:bt_vs_cMB}
Figure \ref{fig:Moni_MW} shows the disk surface density derived
by using the BT estimator ($\Sigma_{BT}$) and by using the cMB estimator 
($\Sigma_{cMB}$). To facilitate comparison with Moni Bidin et al.~(2015), 
we assume $V_{c,0}=215$ km s$^{-1}$, $\xi=0$ and
use $\sigma_{Rz,1}^2$, as given in Eq.~(\ref{eq:sigmaRz1}).
As already stated in Section \ref{sec:data}, there are no 
data for $\partial \bar{V}_{\phi}/\partial R$ beyond $z=2.7$ kpc. Therefore,
the inferences at $z>2.7$ kpc, using the extrapolation of 
$\partial \bar{V}_{\phi}/\partial R$, are shown just to compare the
predictions of the different estimators when the same value for 
$\partial \bar{V}_{\phi}/\partial R$ is used in all cases. 

The integrals in Equation (\ref{eqn:mb14_correct}) have lower limit $z=0$.
Since we do not have measurements of the velocity dispersions
of our tracer population below a height of $1.5$ kpc, Equations 
(\ref{eq:sigmaR})-(\ref{eq:sigmaRz1}) were
extrapolated down to $z=0$.
The contribution of the term given in Equation (\ref{eq:missingterm})
cannot be ignored in general. For instance, for 
$h_{z}=0.9$ kpc and $h_{\sigma_{Rz}}=3$ kpc, its contribution to
$\Sigma_{cMB}$ is $-44.6 M_{\odot}$ pc$^{-2}$ at $z=4$ kpc.

In order to show the sensitivity of the surface density 
estimators to changes in the geometrical parameters of the thick disk, we fix
$h_{z}$ to $0.9$ kpc and explore different combinations of $h_{R}$ 
and $h_{\sigma}$, where we assumed that $h_{\sigma_{R}}=h_{\sigma_{\phi}}=
h_{\sigma_{Rz}}\equiv h_{\sigma}$. 
Figure \ref{fig:Moni_MW} indicates that $\Sigma_{cMB}$ 
is more sensitive to changes in the geometrical parameters than
$\Sigma_{BT}$. Whereas
$\Sigma_{BT}(4\,{\rm kpc})$ varies between $95M_{\odot}$ pc$^{-2}$
to $130M_{\odot}$ pc$^{-2}$ for the combinations of parameters explored,
$\Sigma_{cMB}(4\,{\rm kpc})$ varies between $0M_{\odot}$ pc$^{-2}$
to $110M_{\odot}$ pc$^{-2}$.
Indeed, $\Sigma_{cMB}$ strongly depends on $h_{\sigma}$. 
The most sensitive coefficient to $h_{\sigma}$ is $k_{1}$; for instance,
for $h_{R}=3.5$ kpc, $k_{1}$ is reduced by a factor $7.6$ when 
$h_{\sigma}$ is varied from $3$ kpc to $5$ kpc.
The strong dependence of $\Sigma_{cMB}(z)$ on $h_{\sigma}$ immediately
indicates that we need a good measure of $h_{\sigma}$ for the cMB
estimator to be useful.  If $h_{\sigma}$ is fixed to $5$ kpc,
the cMB estimator is very robust to changes in $h_{R}$ and
it holds that $\Sigma_{BT}(z)\lesssim \Sigma_{cMB}(z)$.

 Moni Bidin et al.~(2012b) noticed that,
assuming $h_{\sigma}=h_{R}$, their surface density estimations
increased with $h_{R}$ and concluded that 
in order to have $\Sigma (4{\,{\rm kpc}})\gtrsim 100 M_{\odot}$pc$^{-2}$,
as extrapolated from the Galactic rotation curve,
the thick disk should have an excessively large scalelength $h_{R}$ 
($h_{R}\gtrsim 4.7$ kpc).
Figure \ref{fig:Moni_MW} shows that the relevant scale is not $h_{R}$
but $h_{\sigma}$. Indeed, for $h_{\sigma}=5$ kpc,
we have $\Sigma_{cMB}({4\,{\rm kpc}})\simeq 105 M_{\odot}$pc$^{-2}$,
even if the radial scalelength $h_{R}$ is relatively short 
($h_{R}\simeq 2$ kpc).

In some cases, especifically for $h_{R}\leq 2$ kpc and $h_{\sigma}\leq 3$ kpc, 
the cMB estimator provides a (unphysical) declining estimate for 
$\Sigma$ with $z$, when using the linear fit given in
Eq.~(\ref{eq:dVphi_dR}) for $\partial \bar{V}_{\phi}/\partial R$.  
If $\partial \bar{V}_{\phi}/\partial R$
is larger than predicted by our linear fit, the declining trend
of $\Sigma_{cMB}$ with $z$, found for those cases, can be
alleviated. To illustrate how $\Sigma_{cMB}$ depends
on $\partial \bar{V}_{\phi}/\partial R$, we also show $\Sigma_{cMB}$ 
when using a power-law fit of the form 
$\partial \bar{V}_{\phi}/\partial R=4|z|^{1.5}+1.4$, which also
provides a good fit to the data.

\begin{figure*} %
\epsscale{0.7}
\plotone{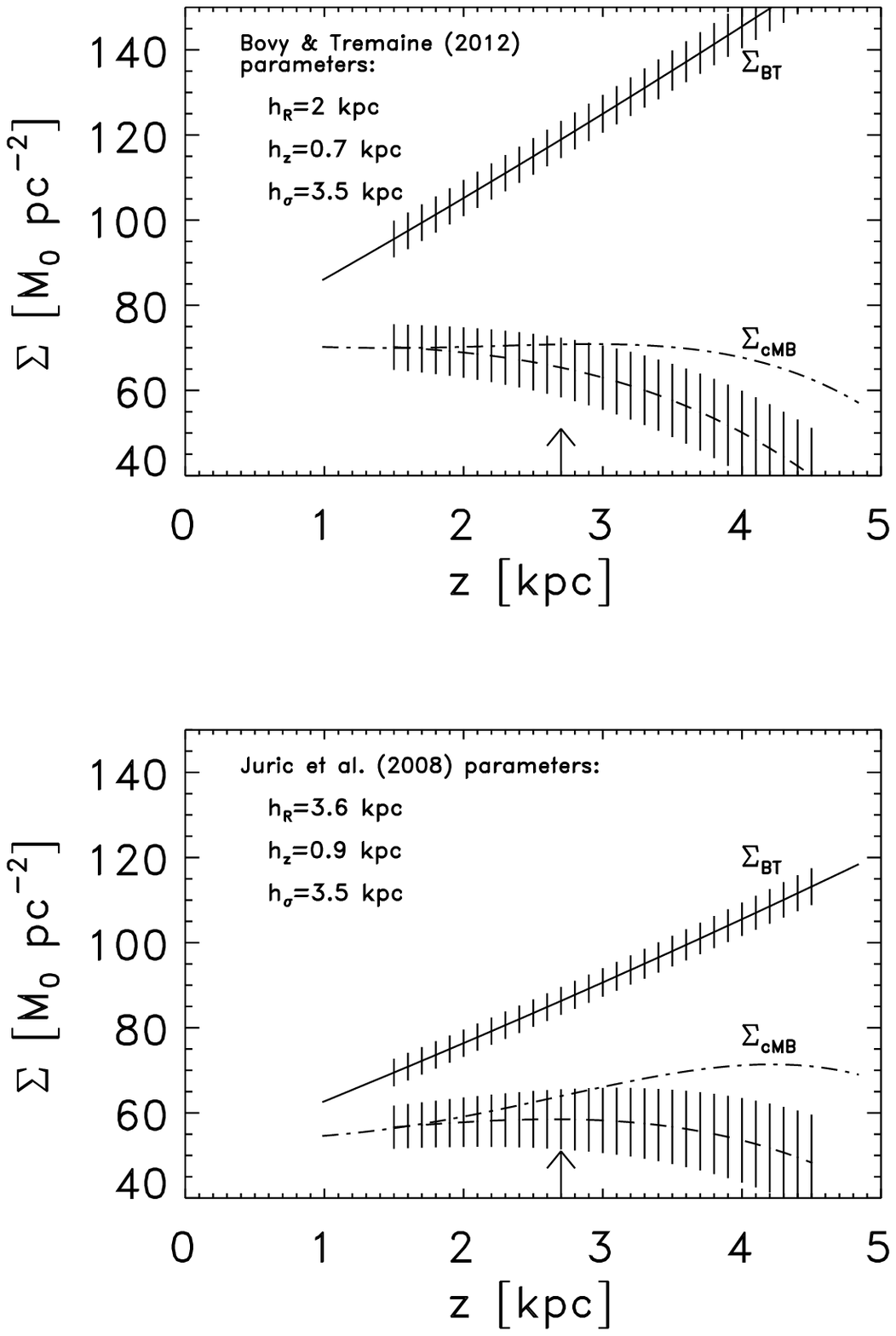}
\caption{Same as Figure \ref{fig:Moni_MW} but using  the parameters of the thick disk preferred by
Bovy \& Tremaine (2012) --upper panel-- and those derived
by Juri\'c et al.~(2008) --lower panel--. 
  }
\label{fig:Moni_MW_param}
\end{figure*}

\begin{figure} %
\epsscale{1.3}
\plotone{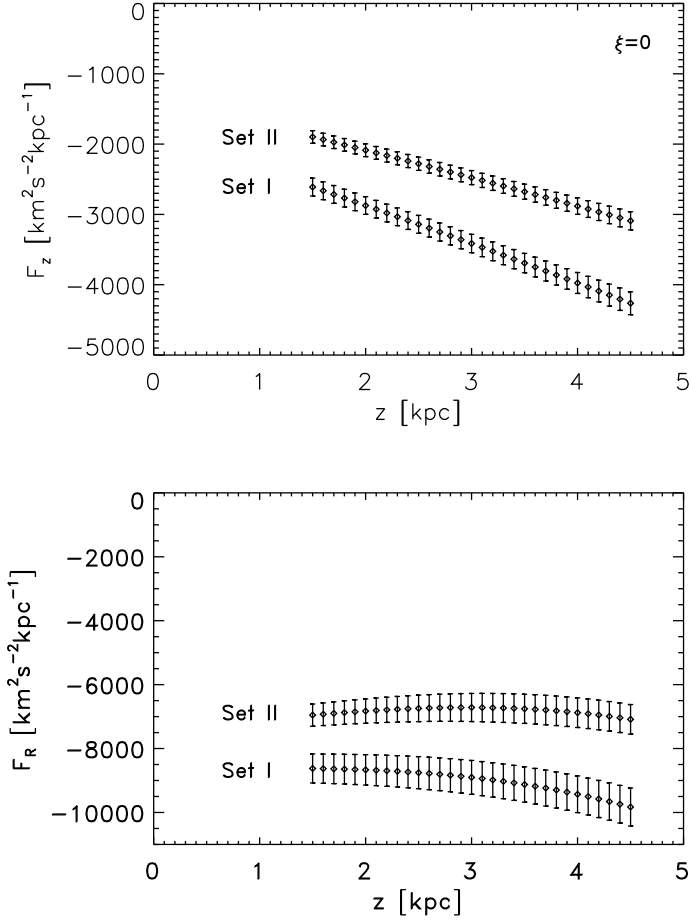}
\caption{Kinematical measurements of the vertical (upper panel) and
horizontal (lower panel) components of the gravitational force at 
$R_{\odot}$, for set I and set II.
}
\label{fig:estimates}
\end{figure}

\begin{figure} %
\epsscale{1.3}
\plotone{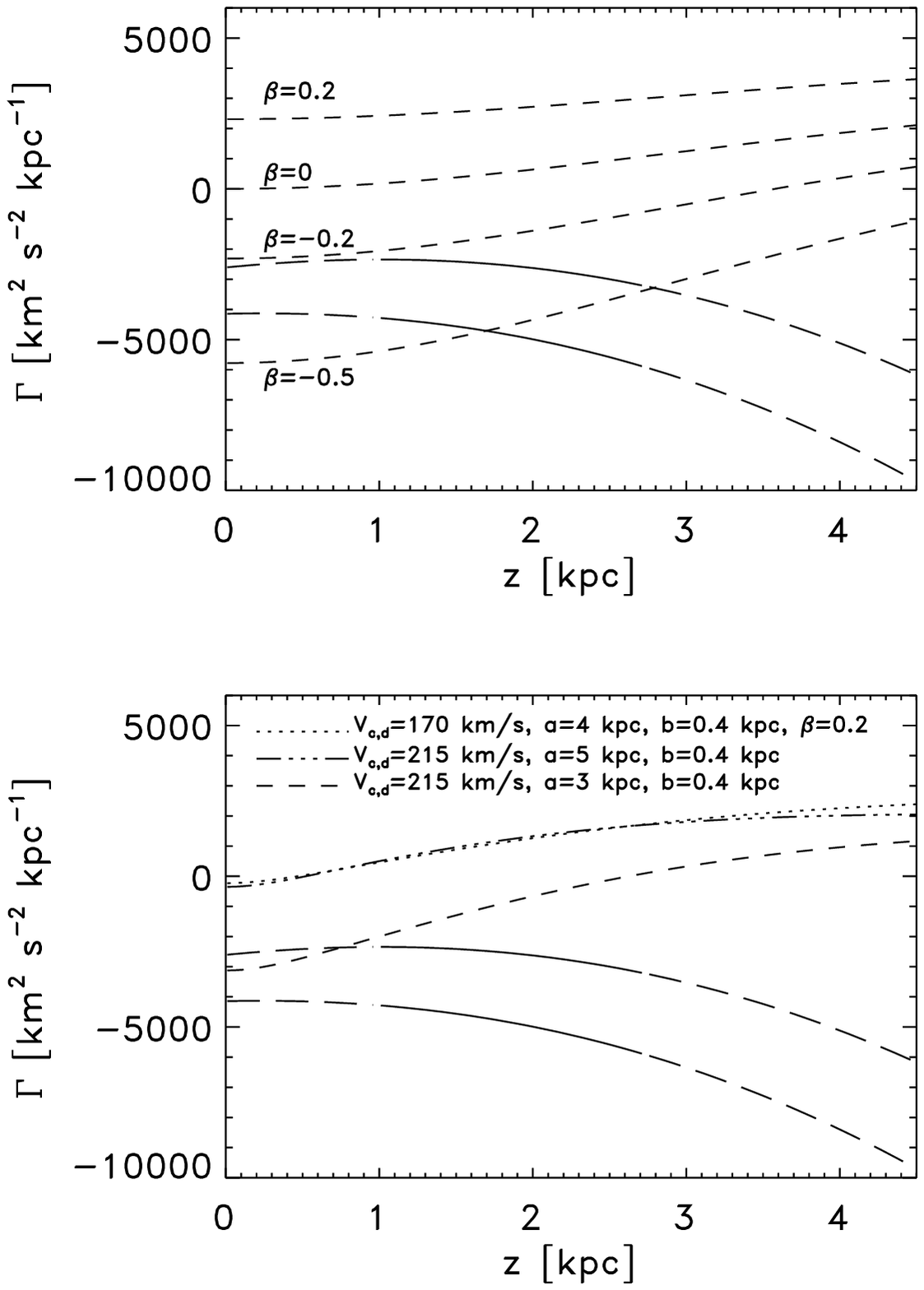}
\caption{
Dependence of $\Gamma$ (the radial derivative of $V_{c}^{2}$), 
as a function of height.
The solid lines represent the Milky Way estimated values using 
Equation (\ref{eq:dVc_dR}) for two sets of parameters: set I
(lower solid curve in each panel) and set II
(upper solid curve in each panel). 
The long dashed line indicates that we are using
an extrapolation of the linear fit of 
$\partial \bar{V}_{\phi}/\partial R$, which
was inferred from data at $0.6$ kpc$<z<2.7$ kpc. Also shown are 
$\Gamma$ for several different mass models consisting of a 
spheroidal component only (upper
panel) or a spheroidal component plus a Miyamoto-Nagai disk with parameters $a$ and $b$
(lower panel).}
\label{fig:dVc2_dR}
\end{figure}

Figure \ref{fig:Moni_MW_param} shows the estimates of the surface density for
two sets of the geometrical parameters. The first set, named
as set I, corresponds to $h_{R}=2$ kpc, $h_{z}=0.7$ kpc and 
$h_{\sigma}=3.5$ kpc, and it is the set preferred by Bovy \& Tremaine (2012).
The second set of parameters are $h_{R}=3.6$ kpc, $h_{z}=0.9$ kpc and
$h_{\sigma}=3.5$ kpc, and were derived by Juri\'c et al.~(2008).
The differences between $\Sigma_{BT}$ and $\Sigma_{cMB}$ are remarkable.
The mismatch between the values predicted by BT and cMB
estimators is larger at high $z$ and for the parameter set I.
For both set of geometrical parameters, it holds that 
$\Sigma_{cMB}(z)<\Sigma_{BT}(z)$. In fact, for 
the model assumptions and the data compiled in \S\ref{sec:data}, $S_{F_{R}}(z)<0$,
and, moreover, $S_{F_{R}}$ is comparable in magnitude to $S_{F_{z}}$;
in particular, $|S_{F_{R}}|\simeq 0.5 S_{F_{z}}$ at $z=4$ kpc.

\begin{figure*} %
\epsscale{1.1}
\plotone{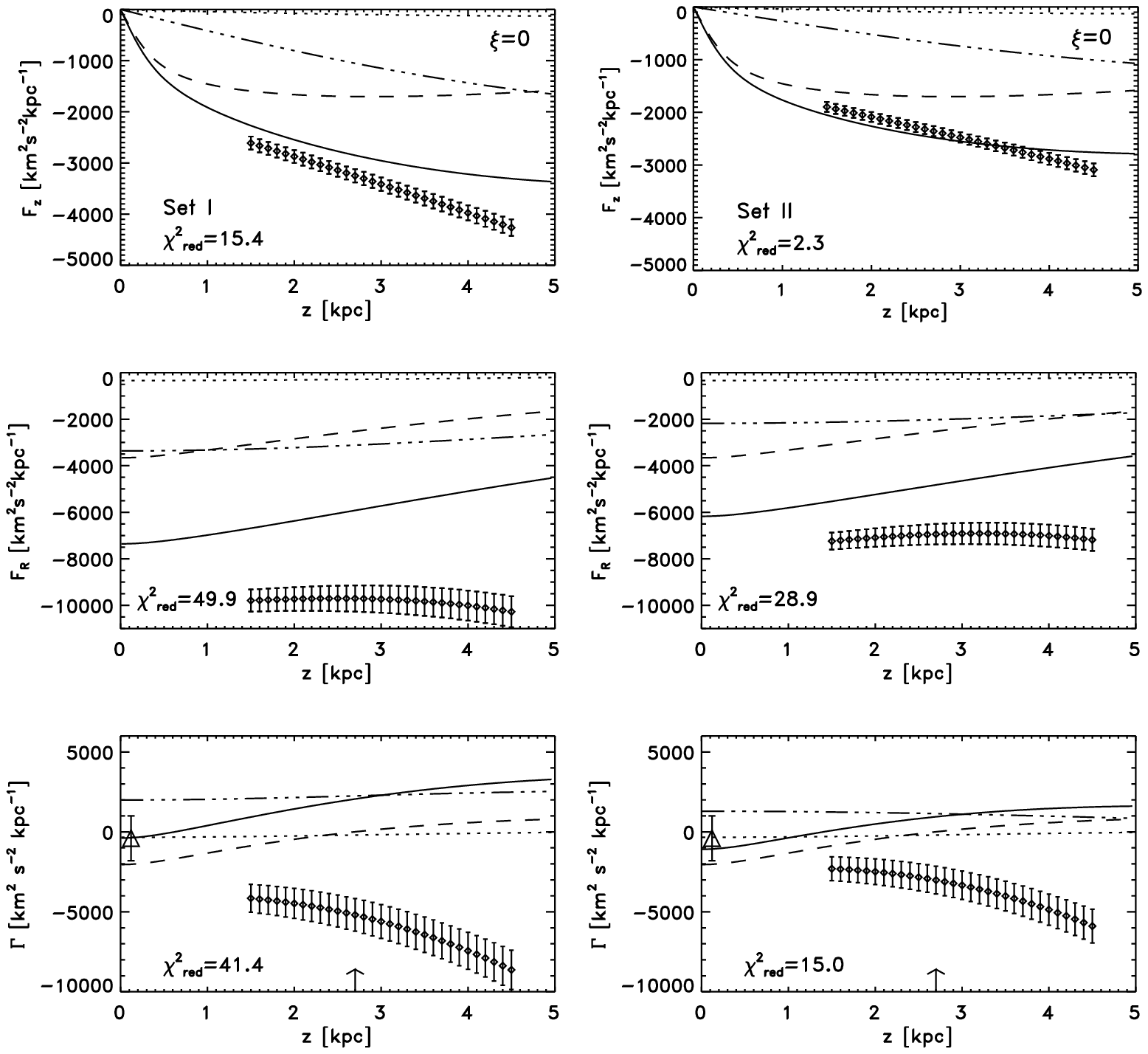}
\caption{Vertical profiles of $F_{z}$,
$F_{R}$ and $\Gamma$, at $R_{\odot}$. 
The kinematical estimates were derived
using Eqs.~(\ref{eq:fzest}), (\ref{eq:fRest}) and (\ref{eq:dVc_dR})
(dots with error bars) with the parameters set I (left
column) and with set II (right column).
The solid line represent the values in the mass model A (left column)
and in mass model B (right column).
In these models, the DM resides in an NFW halo.
For the baryonic mass, we used the model described in 
\S \ref{sec:parametricandmodel}.
The contribution of each component
is also shown: the dark component (triple dot-dashed line),
visible disk (dashed line) and bulge (dotted line).
The triangle with the bar indicates the interval of values derived using
recent determinations of the Oort constants (see Table \ref{table:parameters}).
The arrows at the bottom panels remind the reader that we have
extrapolated the observed values of $\partial \bar{V}_{\phi}/
\partial R$ at heights $>2.7$ kpc. The quoted
value of $\chi^{2}_{\rm red}$ for $\Gamma$
was computed within $1.5<z<2.7$ kpc.
}
\label{fig:BT_NFW}
\end{figure*}

It can be seen that $\Sigma_{BT}(z)$ increases linearly with $z$, implying
that the density of the DM halo is approximately constant from $z=1.5$ kpc
to $4$ kpc.  The mean volume density of DM between $z=1.5$ kpc and $z=4$ kpc
predicted by the BT estimator is $\bar{\rho}_{\rm dm}\simeq 0.010M_{\odot}$
pc$^{-3}$ for set I, and $\bar{\rho}_{\rm dm}\simeq 0.007M_{\odot}$pc$^{-3}$ 
for set II.

On the contrary, $\Sigma_{cMB}(z)$ hardly increases 
from $z=1$ to $z=3$ kpc (see also Moni Bidin et al.~2015), which 
would indicate that the DM density at $|z|>1$ kpc is small.
In fact, since $\rho(z)=(1/2)d\Sigma/dz$, one may conclude 
that, if $\Sigma_{cMB}$ is a good estimator and $h_{\sigma}\leq 3.5$ kpc,
then the DM density at $|z|>1$ kpc is negligible for the parameter set I.
Even adopting the most favorable assumptions (parameter set II and the
power-law fit for $\partial \bar{V}_{\phi}/\partial R$), the surface
density increases $10M_{\odot}$pc$^{-2}$ from $z=1$ kpc to $z=2.5$ kpc,
leading to $\bar{\rho}_{\rm dm}=0.002\pm 0.0025M_{\odot}$pc$^{-3}$ at 
$|z|>1$ kpc.  The same value was obtained by Moni Bidin et al.~(2015). 
In the next Section, we explore the differences between the
predictions of $\Sigma_{BT}$ and $\Sigma_{cMB}$ in more detail.

\subsection{The gravitational force components and $\Gamma$}
\label{sec:gravforcescomp}
Figure \ref{fig:estimates} shows $F_{z}^{\rm est}$
and $F_{R}^{\rm est}$ for the parameter sets I and II, 
using Eqs.~(\ref{eq:fzest}) and (\ref{eq:fRest}). 
It can be seen that the absolute value of $F_{z}^{\rm est}$ increases
almost linearly with $z$ (that is the reason for the almost linear
increase of $\Sigma_{BT}$ with $z$). The vertical profile 
of $|F_{R}^{\rm est}|$, 
on the other hand, is almost flat, with a slight trend to
increase with $z$, particularly for set I. 
This behaviour of $F_{R}^{\rm est}$ with $z$ is against expectations
because oblate mass distributions predict that $|F_{R}|$ must decrease
with $z$. This incorrect trend suggests that $F_{R}^{\rm est}$ is
contaminated by systematic uncertainties or poor modeling.
Moreover, the values of $|F_{R}^{\rm est}|$ are also larger than
expected; for the adopted circular velocity of $215$ km s$^{-1}$, 
$F_{R}$ at $z=0$ should be 
$-V_{c,0}^2/R_{\odot}=-215^{2}/8=-5800$ km$^{2}$ s$^{-2}$ kpc$^{-1}$.
Since $|F_{R}|$ should decrease with $z$, then $|F_{R}|\leq 5800$ 
km$^{2}$ s$^{-2}$ kpc$^{-1}$ at any $z$. However, $|F_{R}^{\rm est}|>8500$ 
km$^{2}$ s$^{-2}$ kpc$^{-1}$ for set I and $|F_{R}^{\rm est}|>6500$
km$^{2}$ s$^{-2}$ kpc$^{-1}$ for set II.
This also suggests that the adopted value for $V_{c,0}$ is not consistent
with the data used to derive $F_{R}^{\rm est}$, especially for set I.

\begin{figure*} %
\epsscale{1.1}
\plotone{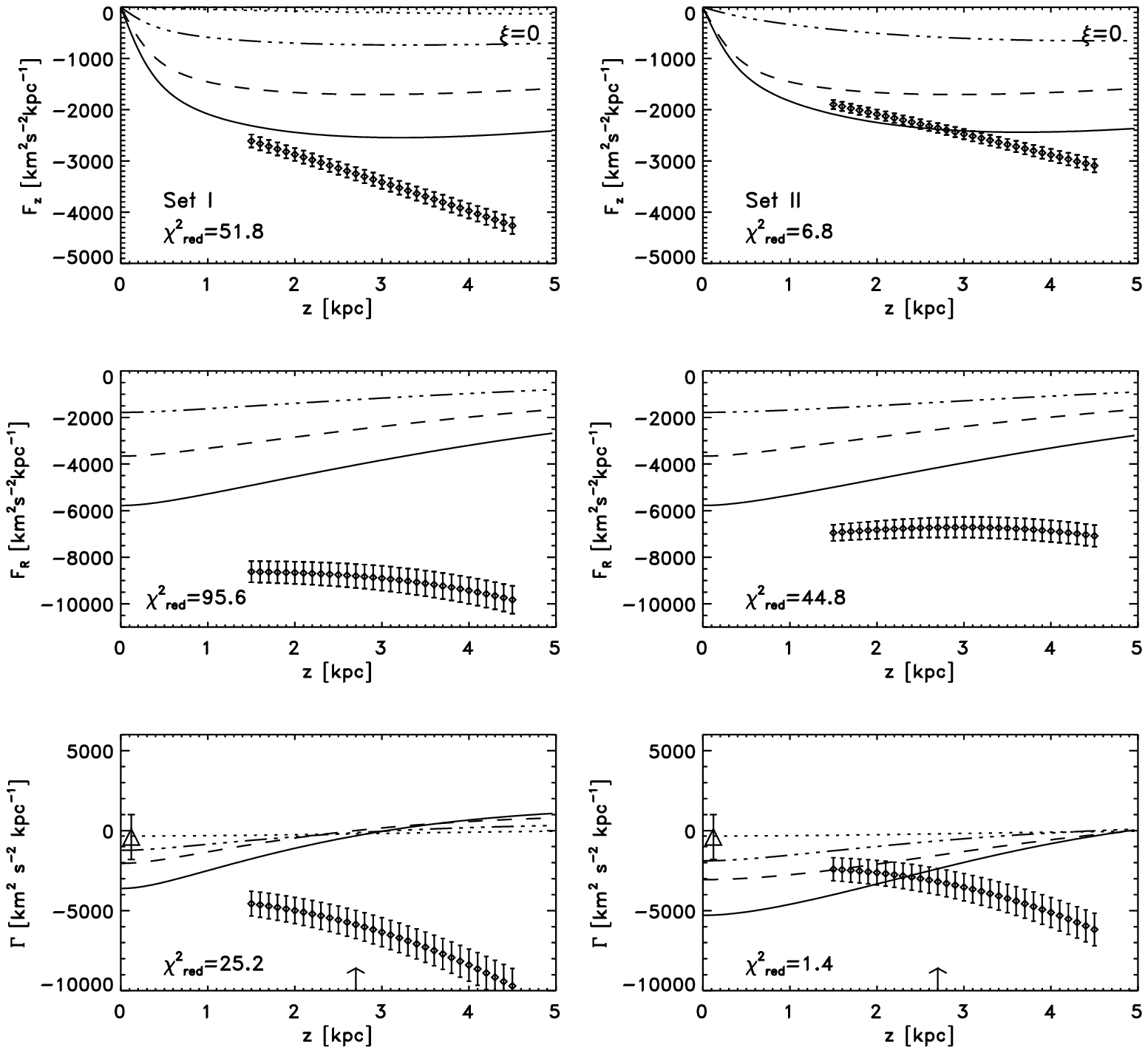}
\caption{
Same as Figure \ref{fig:BT_NFW} but for models C (left column)
and D (right column). In these models  
the DM resides in a thin exponential disk. 
The radial scale lengths of the DM disk
were chosen in order to have a local circular velocity of $215$ km s$^{-1}$.
}
\label{fig:Moni_nodark}
\end{figure*}

The vertical profile for $\Gamma$ can be derived kinematically
using Equation (\ref{eq:dVc_dR}),
with $\sigma_{ij}$, $\bar{V}_{\phi}$ and $\partial \bar{V}_{\phi}/\partial R$
given in \S \ref{sec:data}. 
Note that $S_{F_{R}}$ is simply the integral of 
$\Gamma/R$ over $z$ (see Equation \ref{eq:sfR}).
Figure \ref{fig:dVc2_dR} shows that $\Gamma^{\rm est}$ 
decreases with $z$ (i.e.~$\partial \Gamma^{\rm est}/\partial z<0$), 
acquiring large negative values at $z=4$ kpc. 
In order to show that this trend for $\Gamma$ is
unphysical, we overplot $\Gamma$ as 
a function of $z$ at $R=R_{\odot}=8$ kpc, for theoretical
mass models consisting of a spherically-symmetric component
(representing the halo and the bulge), plus a 
Niyamoto-Nagai (1975) disk with scalelengths $a$ and $b$. 
The spherical component is characterized by the $\beta$-parameter
defined as $\beta\equiv d\ln V_{\rm tilt}/d\ln r$,
where $V_{\rm tilt}^{2}\equiv r d\Phi_{sph}/dr$ and $\Phi_{sph}$ is
the potential associated with the spherical component.
The expressions for the $R$-derivative of $V_{c}^{2}$ are given 
in the Appendix \ref{sec:introproblems}.

\begin{table*}
\begin{minipage}{90mm}
\caption[]{Summary of the representative mass models}
\vspace{0.01cm}

\begin{tabular}{c c c c c c }\hline
{Model} & {Dark matter} &{$\bar{\rho}_{\rm dm}$}& {$\Sigma (1.5\,{\rm kpc})$} &
{$\Sigma (4\,{\rm kpc})$} & {$V_{c,0}$} \\
 & profile &{M$_{\odot}$pc$^{-3}$}& {M$_{\odot}$pc$^{-2}$} &
{M$_{\odot}$pc$^{-2}$} & {km s$^{-1}$} \\
\hline
A & NFW & 0.0116 & 85 & 143 & 243 \\
B & NFW & 0.0078 & 71 & 110 & 222 \\
C & Thin exp.~disk & $\simeq 0.0$ & 70 & 70 & 215 \\
D & Thin exp.~disk & $\simeq 0.0$ & 57 & 57 & 215 \\
\hline
\end{tabular}
\label{table:models}
\end{minipage}
\end{table*}

We explore different relative mass contributions of the disk but
in all our models, we have assumed that the circular velocity
in the midplane is $215$ km s$^{-1}$, at $R_{\odot}=8$ kpc.
We see that the slope of $\Gamma^{\rm est}$ vs $z$ 
cannot be accounted for with a disk and a spherical component;
in all the mass models $\partial\Gamma/\partial z>0$.
In spherical models, $\Gamma\geq 0$ 
at $z>z_{\rm turn}$, where $z_{\rm turn}\leq R/\sqrt{2}$ (see
Appendix \ref{sec:introproblems}).
In models having a pure disk, $\Gamma>0$ if $a\geq 4$
kpc.
Negative values for $\Gamma$, as derived 
from the data (especially for the parameter set I), are very difficult to 
account for at $z>2$ kpc.
In fact, values below $-5000$ km$^{2}$ s$^{-2}$ kpc$^{-1}$ 
at $z>2$ kpc, as those predicted using the geometrical parameter
set I, are not possible
even with a very compact mass distribution where all mass were located 
interior to $R_{\odot}$ (case $\beta=-0.5$).

Since the derived values of $\Gamma^{\rm est}$
are more negative than allowed from generic mass distributions, our
first conclusion is that $|S_{F_{R}}|$ is probably 
being overestimated and thereby $\Sigma_{cMB}$ underestimates the
surface density in the cases under consideration. 
It is possible that $|\Gamma^{\rm est}|$ is being
overestimated because (1) either $V_{c,0}$ or $\partial \bar{V}_{\phi}/\partial R$,
or both, have been
underestimated, (2) there are heuristics involved in computing
the uncertainties in the observed quantities (e.g., Sanders 2012), 
(3) the mass distribution of the tracer stars cannot be described
with a double exponential density profile,
(4) the assumptions that the squared velocity dispersions
vary all with the same scale length
$h_{\sigma_{R}}=h_{\sigma_{\phi}}=h_{\sigma_{Rz}}=3.5$ kpc and that these
scale lengths do not depend on $z$,
are not good approximations, (5) the flaring of the thick disk must be 
taken into account or (6) a combination of them. Regarding point (4), 
we comment that measurements of the radial scale of the
velocity dispersion in the Milky Way are scarce.
Lewis \& Freeman (1989) measured the radial variation of $\sigma_{R}^{2}$
and $\sigma_{\phi}^{2}$ in the plane of the old disk from about $1$
to $17$ kpc in galactocentric radius and found 
$h_{\sigma_{R}}=4.37\pm 0.32$ kpc and $h_{\sigma_{\phi}}=3.36\pm 0.62$ kpc.
The difference between $h_{\sigma_{R}}$ and $h_{\sigma_{\phi}}$
was interpreted as suggesting that the rotation curve is not precisely
flat in the range of galactocentric distances covered by observations.
On the other hand, the radial scale of $\sigma_{z}$ was found
to be $3.5$ kpc by (Bovy et al.~2012a) and $4.1\pm 1$ kpc
by Hattori \& Gilmore (2015). Thefore, there is 
no reason to assume that all the components of the velocity dispersion 
tensor have the same scale lengths. 

In an attempt to assess the reliability of the dynamical estimates
of the local surface density and to check if their unexpected too low values
were an artifact of the assumed kinematics, Moni Bidin et al.~(2012b)
repeated the calculations using two independent data sets for the
kinematics of the thick disk stars. It is remarkable that the
three sets returned similar values for the surface density. This
might indicate that either all data sets are affected by the
same systematics, or the model assumptions are not adequate (or both). 
In the next Sections, we will try to shed light on these issues.

\subsection{Representative mass models}
\label{sec:representative}
In the previous subsection, we showed that both $F_{R}^{\rm est}$ and
$\Gamma^{\rm est}$ take values more negative than expected
if $V_{c,0}=215$ km s$^{-1}$. It is worthwhile to calculate
$F_{z}$, $F_{R}$ and $\Gamma$ in  
simple global mass models and compare them with the values derived as
using the kinematical data of the thick disk. To do so, we consider
four representative mass models, referred as models A, B, C and D.
A summary of the relevant parameters of each model is given in 
Table \ref{table:models}. We also give $\bar{\rho}_{\rm dm}$ which represents
the mean volume density of DM between $z=1.5$ kpc and $z=4$ kpc.
All the mass models have the same distribution of baryonic mass,
but differ in the dark matter distribution.  
The baryonic mass is distributed as follows (see \S\ref{sec:baryonicmodel} 
for details and references); a bulge of mass $5\times 10^{9}M_{\odot}$, 
a radially exponential stellar disk with scalelength $2.15$ kpc,
and a local surface density of $38M_{\odot}$pc$^{-2}$, 
the ISM layer modeled as a radially exponential disk with a scalelength 
of $8.5$ kpc and a local surface density of $13M_{\odot}$pc$^{-2}$.

Models A and B have local surface densities $\Sigma(4\,{\rm kpc})$     
similar to the values derived using the BT estimator 
for set I and set II, respectively
(see \S \ref{sec:bt_vs_cMB} and Table \ref{table:models}).
For these models, we assume the DM is distributed in a NFW halo with
a scale radius of $19$ kpc. For this scale radius, the circular
velocity at $R_{\odot}$ is $\sim 220-245$ km s$^{-1}$ (see Table
\ref{table:models}).
In particular, the parameters of the DM halo in model A are
similar to those inferred by Nesti \& Salucci (2013) to fit the
rotation curve of our Galaxy with a NFW profile.

In \S\ref{sec:bt_vs_cMB}, we found that $\Sigma_{cMB}$ predicts a 
surface density of $\simeq 70M_{\odot}$pc$^{-2}$ at $z=1.5$ kpc for the set I,
and a surface density of $\simeq 57M_{\odot}$pc$^{-2}$ for the set II.
Thus, assuming that the surface density in visible matter is 
$\simeq 51M_{\odot}$pc$^{-2}$ (eg., Bovy \& Rix 2013), 
$\Sigma_{cMB}$ would imply that the DM should
lie in a thin disk with a scaleheight $\lesssim 1$ kpc and a local
surface density $\sim 19M_{\odot}$pc$^{-2}$ (parameters set I)
and $\sim 6M_{\odot}$pc$^{-2}$ (set II).
To mimic these density profiles,
models C and D assume that there is a DM component, which is deposited in 
a radially exponential disk with a scaleheight $<1$ kpc and
local surface density of $19M_{\odot}$pc$^{-2}$ (model C) and
$6M_{\odot}$pc$^{-2}$ (model D).
The radial scalelength for the DM disk was chosen in order to 
give a local circular velocity of $215$ km s$^{-1}$, to be consistent
with the value adopted in \S\ref{sec:bt_vs_cMB}.
More specifically, the radial scalelength for the DM disk is $2.1$ kpc 
for model C and $1.4$ kpc for model D.

In Figures \ref{fig:BT_NFW} and \ref{fig:Moni_nodark}, we compare 
the values of $F_{z}$, $F_{R}$ and $\Gamma$
of the models A, B, C and D with the kinematical measurements of
these quantities.
In order to compare the different models, the reduced chi-square,
$\chi^{2}_{\rm red}$, defined as $\chi^{2}/N$, where $N$ is the
number of data points,
are given at each panel. In the case of $\Gamma$,
the quoted value of $\chi^{2}_{\rm red}$ was calculated using only
the data points within $1.5$ kpc $\leq z\leq 2.7$ kpc because there
exist direct measurements of $\partial \bar{V}_{\phi}/\partial R$
only up to $z=2.7$ kpc (see \S\ref{sec:data}).

The linear shape of $F_{z}^{\rm est}$ cannot be reproduced if all
the mass is distributed in a disk (models C and D; 
see upper panels in Fig.~\ref{fig:Moni_nodark}).
As expected from the analysis in \S\ref{sec:gravforcescomp}, all the 
mass models have $|F_{R}|$ values that are smaller
than $|F_{R}^{\rm est}|$. For instance, $|F_{R}^{\rm est}|$ at $z=4$ kpc 
(calculated using the parameters set I) is a factor of $2-3$
larger than it is in the mass models A and C. 
The fact that both $|F_{z}^{\rm est}|$ and $|F_{R}^{\rm est}|$
are larger than they are in the mass models A and C, suggests that 
the data allows a more massive dark component than adopted in these mass
models.

On the other hand, the profile of $\Gamma$,
as derived from the kinematics of the thick disk, is much more negative
than in the mass models. The discrepancies in $F_{R}$ and also in 
$\Gamma$ are somewhat mitigated when the parameters
set II are used. 

From the $\chi^{2}$-statistics, there is no any clear preference for model B
or for model D. Model B explains better $F_{z}^{\rm est}$
and $F_{R}^{\rm est}$ but model D is better in accounting for the large 
negative values of $\Gamma^{\rm est}$ between $z=1.5$ kpc and $z=2.7$ kpc.
A remarkable difference between models B and D is the value of $\Gamma_{0}$,
defined as $\Gamma_{0}\equiv \Gamma(R_{\odot},0)$.
Model B has $\Gamma_{0}=-1000$ km$^{2}$ s$^{-2}$ kpc$^{-1}$, whereas
Model D has $\Gamma_{0}=-5200$ km$^{2}$ s$^{-2}$
kpc$^{-1}$ (see solid lines in the lowest right panels
in Figures \ref{fig:BT_NFW} and \ref{fig:Moni_nodark}).  
Good observational measurements of $\Gamma_{0}$
at $z=0$ are crucial to discriminate between different models; indeed,
very different mass models may have similar values of both $F_{z}$
and $F_{R}$ at $0\leq z\leq 4$ kpc. 

\begin{table*}
\begin{minipage}{106mm}
\caption[]{Estimates of the $R$-derivative of $V_{c}^{2}$, $\Gamma$, at $z=0$
 }
\vspace{0.01cm}

\begin{tabular}{c c c c c}\hline
{Stellar} & {$A$} &{$B$}& {$\Gamma_{0}$}& {References} \\
{Type}& {km s$^{-1}$kpc$^{-1}$}& {km s$^{-1}$kpc$^{-1}$} &  {{km$^{2}$ s$^{-2}$kpc$^{-1}$}}& {} \\
\hline
Cepheids & $14.82\pm 0.84$ & $-12.37\pm 0.64$ & $ -1100 \pm 500$ & Feast \& Whitelock (1997) \\
K, M giants & $14.5\pm 1.0$ & $-11.5\pm 1.0$ & $-1250\pm 600$ & Mignard (2000) \\
K, M giants & $15.86\pm 1.30$ & $-14.57 \pm 1.01$ & $-650\pm 800$ & Yuan et al.~(2008) \\
Red giants & $9.6\pm 0.5$ & $-11.6\pm 0.5 $ & $700\pm 250$ & Olling \& Dehnen (2003) \\
F giants & $14.85\pm 7.47$ & $-10.85\pm 6.83 $ & $-1650\pm 4000$ & Branham (2010) \\
APOGEE  & $13.5^{+0.2}_{-1.0}$ & $-13.7^{+3.3}_{-0.1} $ & $100^{+100}_{-1300}$
& Bovy et al.~(2012b) \\

\hline
\end{tabular}
\label{table:parameters}
\end{minipage}
\end{table*}

It is generally assumed that the inner rotation curve is rather flat,
typically $\partial \ln V_{c}/\partial \ln R=0\pm 0.15$ at $z=0$ 
(e.g., Salucci et al.~2010).
The most common way to estimate $\Gamma_{0}$ is via the Oort constants $A$ and $B$:
\begin{equation}
\Gamma_{0}=2R(B^{2}-A^{2}).
\end{equation} 
For recent values of the Oort constants derived 
using the proper motion of relatively old giant stars, 
which are one of the most reliable tracers, we find that 
$\Gamma_{0}$ lies between $-2000$ and $1000$ km$^{2}$
s$^{-2}$kpc$^{-1}$ (see Table \ref{table:parameters}).
From the velocity distribution of stars in the thin-disk population,
Fuchs et al.~(2009) found  
that the rotation curve at $R_{\odot}$ is almost flat
($\partial \ln V_{c}/\partial \ln R=-0.006\pm 0.016$ at $z=0$). 
Therefore, model B is more in accordance with the observational inferences
of $\Gamma_{0}$. 

We have also explored other variants of models C and D.
For instance, 
exponential dark disks having the same local surface
density, but a larger radial scalelength 
($8.5$ kpc). In these models, a spherical DM component
with null density at $r>R_{\odot}$, was added to satisfy the condition
that $V_{c,0}=215$ km s$^{-1}$. 
No significant changes in $F_{z}$, $F_{R}$ or $\Gamma$ were found
in these models as compared to models C and D.

\begin{figure} %
\epsscale{1.16}
\plotone{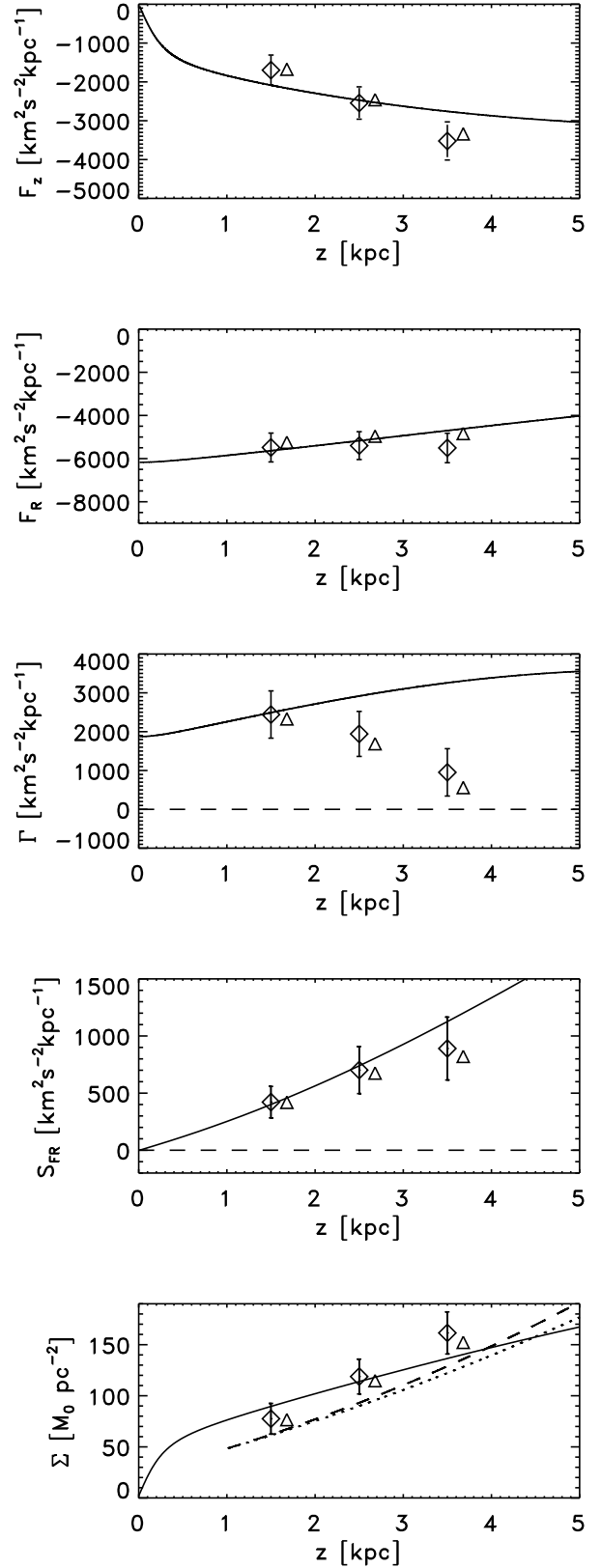}
\caption{Comparison of the exact values of $F_{z}$, $F_{R}$, 
$\Gamma$, $S_{F_{R}}$ and $\Sigma$ in
our simulations (solid lines), as a function of height, with 
their corresponding dynamical measurements, using 
Eqs.~(\ref{eq:fzest})-(\ref{eq:dVc_dR}) with $\xi=0$ (triangles)
and $\xi=-0.02$ (dots with error bars). To make the plot
readable, the triangles have been slightly shifted in the
horizontal direction. The dashed lines 
indicate the values in the BT method.  In the last panel, 
the dashed (dotted) line
represents $\Sigma_{BT}$ for $\xi=-0.02$ ($\xi=0$)
and the dots with error bars are for $\Sigma_{cMB}$.
}
\label{fig:forces_sims}
\end{figure}

\section{Tests using numerical simulations}
\label{sec:sims}
In this Section, we measure different quantities (e.g.,
vertical and radial components of the gravitational force, surface density)
using the kinematics of the tracer population in a simulated galaxy, 
as if it is done for the thick disk of the Milky Way, and then compare 
them with the exact input values.
In particular, we will be able to quantify the role of systematic 
uncertainties and the impact of the assumptions in the above quantites,
particularly, the surface density.

\begin{figure} %
\epsscale{1.16}
\plotone{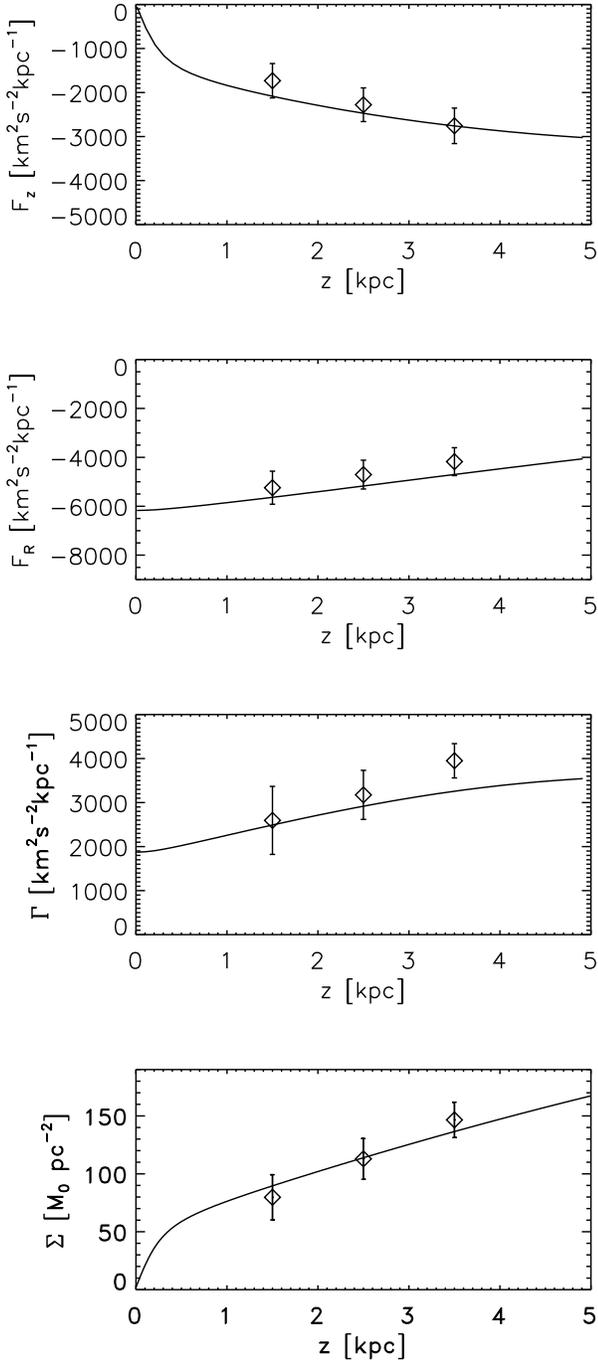}
\caption{Dynamical measurements of $F_{z}$,
$F_{R}$, $\Gamma$ and $\Sigma$ at $R=8$ kpc from our simulations using 
the generalized Equations given in Appendix \ref{sec:second_order}, 
which depend on the local values of the geometrical parameters, 
as they vary with $z$ (dots with error bars).
For comparison, we show also their exact vertical profiles (solid lines). 
}
\label{fig:forces_sims_precise}
\end{figure}

\subsection{Our mock galaxy}
\label{sec:mock_galaxy}
A thick disk-like tracer population, containing $10^{5}$ stars,
was set up in a rigid disk+bulge+dark halo potential. For the gravitational
potential, we adopt the analytical model described in Flynn et al.~(1996).
The model is composed of two spheroidal components (bulge plus a inner core),
three Miyamoto-Nagai disks and a spherical dark halo.
Given the background potential, the exact input
values of $S_{F_{z}}$ and $S_{F_{R}}$ can be calculated.
Since the rotation curve at the midplane in this model is rising at 
the solar position, it holds that $S_{F_{R}}>0$.
At $z>2.5$ kpc, $S_{F_{R}}$ is not negligible as compared to $S_{F_{z}}$.
For instance, $S_{F_{R}}=0.4 S_{F_{z}}$ at $z=3.5$ kpc.
As a consequence, $\Sigma_{BT}$ is expected to underestimate the
surface density by $\sim 28\%$. We wish to test the predictions of 
$\Sigma_{BT}$ and $\Sigma_{cMB}$ in a situation where the tracer population
is a thick disk with a surface density that decays exponentially
with $R$. In the vertical direction, the volume density is
close to exponential as well. 	 

The exponential density scale-height, velocity dispersions and mean 
rotation rates of the tracer stars were set up initially to match, as 
closely as we could, the data of Moni Bidin et al.~(2012a).  
Once the disk was relaxed in the rigid background
potential of the galaxy, the orbits were integrated for $5$ Gyr.

We calculated the velocity dispersions of the stars that are within
$7.5\leq R\leq 8.5$ kpc, at three heights: $z=1.5, 2.5$ and $3.5$ kpc.
At $|z|\geq 1.5$ kpc, the velocity dispersions were fitted well by:
\begin{equation}
\sigma_{R}=(72\pm 3)+(3.2\pm 1)(|z|-2.5)\,\, {\rm km \,\,s^{-1}},
\label{eq:sigmaRsims}
\end{equation}
\begin{equation}
\sigma_{\phi}=(61\pm 3)+(2.7\pm 1)(|z|-2.5)\,\, {\rm km \,\,s^{-1}},
\label{eq:sigmaphisims}
\end{equation}
\begin{equation}
\sigma_{z}=(45\pm 3)+(6\pm 1)(|z|-2.5)\,\, {\rm km \,\,s^{-1}}.
\label{eq:sigmazsims}
\end{equation}
The antisymmetric velocity dispersion component $\sigma_{Rz}^{2}$ was fitted by
\begin{equation}
\sigma_{Rz}^{2}=(1000\pm 450)+(550\pm 150)(z-2.5)\,\, {\rm km^{2} \,\,s^{-2}}.
\label{eq:sigmaRzsims}
\end{equation}
The latter expression holds for $z>1.5$ kpc.
We note that the values of $\sigma_{Rz}^{2}$ are somewhat smaller in our
simulations than the observed values reported in Moni Bidin et al.~(2010).

The mean azimuthal velocity $\bar{V}_{\phi}$, at $R=8$ kpc, was 
fit between $z=0.5$ to $z=4$ kpc using $\bar{V}_{\phi}=197-20|z|^{1.3}$ km s$^{-1}$ 
(with $z$ in kpc). The fit is good and has residuals of less than
$4$ km s$^{-1}$. It is remarkable that a similar
$z$-dependence for $\bar{V}_{\phi}(z)$ was found for thick disk
stars by Ivezi\'c et al.~(2008) 
from the analysis of SDSS data, and by Moni Bidin et al.~(2012a) 
combining different data sets.
Finally, we also computed $\partial \bar{V}_{\phi}/\partial R$ at $R=8$ kpc,
from the simulations, and found $\partial \bar{V}_{\phi}/\partial R=2.8|z|+5.5$
km s$^{-1}$ kpc$^{-1}$ (with $z$ in kpc). The residuals are less than
$0.6$ km s$^{-1}$ kpc$^{-1}$.

\subsection{Dynamical tests}
\label{sec:dynamicaltests}
The vertical profiles of $F_{z}$, $F_{R}$, $\Gamma$, $\Sigma_{BT}$
and $\Sigma_{cMB}$ can be derived from the kinematics 
of the tracer population using Equations (\ref{eq:fzest})-(\ref{eq:gen}) 
and compare them with the input values.
The $z$-derivatives of the velocity dispersions in Equations
(\ref{eq:fzest})-(\ref{eq:dVc_dR}) and the integrals over $z$ in
Equation (\ref{eqn:mb14_correct}) were performed analytically using
the linear fits given in Equations 
(\ref{eq:sigmaRsims})-(\ref{eq:sigmaRzsims}).
In the computation of $S_{F_{R}}$ and $\Sigma_{cMB}$,
we used $z=0$ as the lower limit of integration in 
Equation (\ref{eqn:mb14_correct}) as we did in \S\ref{sec:bt_vs_cMB}.

In principle, the geometrical parameters of the simulated thick disk can
be measured from our simulations.  However, they are not constant with height, 
but depend on $z$. In order to quantify the impact of the assumption
that the geometrical parameters are constant, we have computed
$F_{z}^{\rm est}$, $F_{R}^{\rm est}$, $\Gamma^{\rm est}$, 
$S_{F_{R}}^{\rm est}$, $\Sigma_{BT}$ and $\Sigma_{cMB}$ using
the mean geometrical parameters.
We have measured the {\it mean} geometrical parameters in our simulations,
which will be denoted by an asterisk,
by computing the velocity dispersion tensor vs $R$
in the range $4<R<12$ kpc, including 
all the stars within $|z|<2.5$ kpc.  We obtained 
$h_{z}^{\ast}=0.77$ kpc, $h_{R}^{\ast}=4.0$ kpc, 
$h_{\sigma_{R}}^{\ast}=5.5$ kpc,
$h_{\sigma_{\phi}}^{\ast}=4.6$ kpc and $h_{\sigma_{Rz}}^{\ast}=4.1$ kpc.  
Figure \ref{fig:forces_sims} shows $F_{z}^{\rm est}$, 
$F_{R}^{\rm est}$, $\Gamma^{\rm est}$, $S_{F_{R}}^{\rm est}$,
$\Sigma_{BT}$ and $\Sigma_{cMB}$, derived using the mean 
geometrical parameters and the {\it exact value} of $V_{c,0}$. 
Although our simulated thick disk exhibits a slight ``antiflaring'' with
$\xi^{\ast}=-0.02$ at $R_{\odot}$, we show results for both $\xi=-0.02$
and $\xi=0$. 

It can be seen that the dynamical measurement of $F_{z}$ overestimates
the modulus of the vertical force beyond $z=2.5$ kpc and underestimates it
below this height. In other words, the slope of $F_{z}$ versus $z$ is not
well reproduced by the kinematical estimates, but the mean value
of $F_{z}$ between $z=1.5$ kpc and $z=3.5$ kpc is estimated correctly.
The effect of using either $\xi=0$ or $\xi=-0.02$ is small.

The $R$-component of the force is well reproduced, particularly for
$\xi=0$. The
shape of $F_{R}^{\rm est}$ vs $z$ is slightly flatter than
the input simulation values for $\xi=-0.02$. 
On the other hand, the dynamical inferences
of $\Gamma$ deviate significantly from the
real input values at $z>2.5$ kpc.  
$S_{F_{R}}$ is underestimated as well, but
since $S_{F_{R}}$ is an integral of $\Gamma$ over $z$, 
$S_{F_{R}}$ presents a smaller fractional error than $\Gamma$. 

As expected, $\Sigma_{BT}$ underestimates the surface density because
of the assumption $S_{F_{R}}=0$. However, the difference between
the real and the predicted value at $z=3.5$ kpc is only $10-15M_{\odot}$
pc$^{-2}$, that is $\sim 8.5\%$ (much less than the expected value
of $28\%$; see Section \ref{sec:mock_galaxy}). 
The reason is that $S_{F_{z}}$ is overestimated and this compensates
the ignored positive contribution of $S_{F_{R}}$. 

$\Sigma_{cMB}$ reproduces the surface density within the uncertainties.
At $z=3.5$ kpc, the value predicted for $\xi=0$ is closer to the
exact value. Both estimators equally overestimate $S_{F_{z}}$, but
$\Sigma_{BT}$ compensates it by adopting $S_{F_{R}}=0$ and
$\Sigma_{cMB}$ does not.
Note that both estimators overpredict the slope of $\Sigma$
with $z$, mainly because 
$F^{\rm est}_{z}$ is steeper than the real profile. 

As already anticipated, the geometrical parameters depend on $z$.
In our particular simulation, we have measured the local
geometrical parameters $\tilde{h}_{z}$, $\tilde{h}_{R}$
and $\tilde{h}_{\sigma_{ij}}$ (see Appendix \ref{sec:second_order}
for their definitions) and found they increase
with $z$, except $h_{R}$. For instance, $\tilde{h}_{z}$ increases from 
$0.76$ kpc at $z=1.5$ kpc to $0.91$ kpc at $z=3.5$ kpc, and
$\tilde{h}_{R}$ decreases from $3.8$ to $3.0$ kpc in the same range. 
The changes of $\tilde{h}_{\sigma_{ij}}$ are more remarkable:
$\tilde{h}_{\sigma_{R}}$ varies from $6.0$ kpc at $z=1.5$ kpc,
to $\gtrsim 25$ kpc at $z=3.5$ kpc ($\tilde{h}_{\sigma_{\phi}}$ varies
by a factor of $2.5$ and $\tilde{h}_{\sigma_{Rz}}$ by a factor of $1.8$).
Figure \ref{fig:forces_sims_precise} shows that when the 
local scalelengths are used
at each $z$ (see Appendix \ref{sec:second_order}), 
all the values are recovered within the uncertainties,
except the value of $\Gamma$ at $z=3.5$ kpc. Note that the error bars
in Figure \ref{fig:forces_sims_precise} do not include uncertainties
in the measured geometrical parameters. In particular, the second-order 
$R$-derivatives involved in the computation of $\Gamma$ may introduce 
significant bias (see Appendix \ref{sec:second_order}).

In our particular thick disk realization, 
$\Sigma_{cMB}$ overestimates $\Sigma$ at $z=3.5$ kpc
because $h_{z}$ increases with $z$. 
 Were $h_{z}$ nearly constant in the region
of interest, $\Sigma_{cMB}$ would have predicted correctly the
surface density. 
The two estimators should be mutually consistent if the selected
stars represent a homogenous population with constant geometrical
parameters and the underlying potential is such that $S_{F_{R}}$
is small as compared to $S_{F_{z}}$.
If the geometrical parameters were constant but $S_{F_{R}}$ cannot
be ignored as compared to $S_{F_{z}}$, $\Sigma_{cMB}$ is a more
reliable estimator.
In the lack of any information about the vertical dependence of the
geometrical parameters, it is not possible to discern what
estimator is more reliable.

So far, we have assumed that we know the mean geometrical parameters 
with enough accuracy. In real life, uncertainties in the mean values
of the geometrical parameters of the thick disk are large.
In order to test the robustness of the dynamical
estimates to these uncertainties,
we have repeated the calculation assuming $h_{\sigma_{R}}=
h_{\sigma_{\phi}}=h_{\sigma_{Rz}}=3.5$ kpc, instead of the exact values. 
In this case, $F_{z}^{\rm est}$ and $F_{R}^{\rm est}$ are almost
unaltered by the new choice of the geometrical parameters. However, the
new parameters clearly underestimates $S_{F_{R}}$, which becomes close to zero;
thus $\Sigma_{cMB}\simeq \Sigma_{BT}$. Therefore, in this case,
both predictors give similar results.

\section{The parametric method} 
\label{sec:parametricandmodel}
\subsection{Description of the method}
\label{sec:baryonicmodel}
According to \S\ref{sec:sims}, if the selected stars represent a 
homogeneous population, that is not formed by mixing of populations with
different scalelengths,
and if the mean geometrical parameters and $V_{c,0}$, $\bar{V}_{\phi}$
and $\partial \bar{V}_{\phi}/\partial R$ are known with good precision,
$S_{F_{R}}$ is recovered within the uncertainties.
However, in \S \ref{sec:gravforcescomp} we found 
that $S_{F_{R}}$ takes implausibly
large negative values when calculated using the currently availble 
kinematical data of the thick disk. 
Therefore, uncertain input parameters, systematic uncertainties or/and 
incomplete modeling are possibly contaminating $S_{F_{R}}$.

An alternative means of estimating the surface density $\Sigma$
is to fit a parameterized galaxy potential to
$F_{z}^{\rm est}$, $F_{R}^{\rm est}$ and $\Gamma^{\rm est}$.
Once the best-fitting parameters are found, $\Sigma(z)$ can be computed.
We will refer to this as the {\it parametric method}. 
This method provides physically meaningful solutions and it also includes
information of the radial force field, which is in the spirit of
three-dimensional approaches.

We use a simplified model for the Milky Way composed by a bulge, a disk 
and a dark halo. 
To reduce the number of free parameters of the model and to
avoid large levels of degeneracy between parameters, we fix the
parameters of the visible matter and allow to vary the parameters
of the dark halo.
At $r\geq R_{\odot}$, the stellar bulge is modeled as a potential
$\Phi_{b}\simeq -GM_{b}/r$, with a mass $M_{b}$ of 
$5\times 10^{9}M_{\odot}$ (e.g, McMillan 2011). For the local
surface density of the stellar disk, we take  $38M_{\odot}$ pc$^{-2}$ 
(Bovy \& Rix 2013).
Bovy \& Rix (2013) presented a dynamical measurement of
the  mass-weighted Galactic disk scale length $R_{d}$, using the kinematics 
of abundance-selected stellar populations from the SEGUE survey. They
found $R_{d}=2.15\pm 0.14$ kpc. Since we wish to explore models
as consistent as possible with the dynamical constraints derived in
Bovy \& Rix (2013), we have fixed the mass-weighted scalelength
of the disk to $R_{d}=2.15$ kpc, and the mass scale height
to $0.37$ kpc (Bovy \& Rix 2013). To derive the gravitational
forces created by the stellar disk, it was approximated with three 
Miyamoto-Nagai potentials with
the expressions in Smith et al.~(2015), which provide a good
approximation for the potential created by an exponential disk.
The mass distribution of the ISM layer was represented by an exponential disk
with a local surface density of $13 M_{\odot}$ pc$^{-2}$ 
(Holmberg \& Flynn 2000),
and a long radial scalelength of $8.5$ kpc to reproduce the
plateau in the gas surface density (e.g., Wolfire et al.~2003).
We set the vertical height of the gaseous disk to $0.35$ kpc, 
but its exact value is not relevant for the present study.

As discussed in \S \ref{sec:gravforcescomp} and \ref{sec:representative},
the currently available kinematic data for the thick disk stars alone, 
are of insufficient quality to uniquely determine the profile of
$\Sigma(z)$, without making further assumptions.
To break down the degeneracy between models,
we here aim at calculating the parameters of the DM halo and
the geometrical parameters of the thick disk that better match the data, with
the simplistic but restrictive assumption that the DM halo is 
spherically-symmetric.
The dark halo was assumed to have a circular velocity following
a power-law with index $\beta_{h}$, $V_{h}\propto
r^{\beta_{h}}$ in the region of interest, that is between
a galactocentric radius $r_{1}=R_{\odot}=8$ kpc and 
$r_{2}=(R_{\odot}^2+z_{\rm max}^{2})^{1/2}\simeq 9.2$ kpc,
for $z_{\rm max}=4.5$ kpc. 
The two parameters that characterize the dark halo are the
corresponding circular velocity at $R_{\odot}$, denoted by $V_{h,\odot}$, 
and the $\beta_{h}$ parameter. 
We note that we do not force the DM halo to match the 
expected DM density
of any classical model; if there is no need for DM interior to $R_{\odot}$
to explain the local kinematics of the thick disk, we should obtain $V_{h,\odot}\simeq 0$. 
A value for $\beta_{h}$ close to the Keplerian value ($-0.5$) 
would imply that the
kinematics of the thick disk is consistent with a zero local density halo.

\begin{figure*} %
\epsscale{1.2}
\plotone{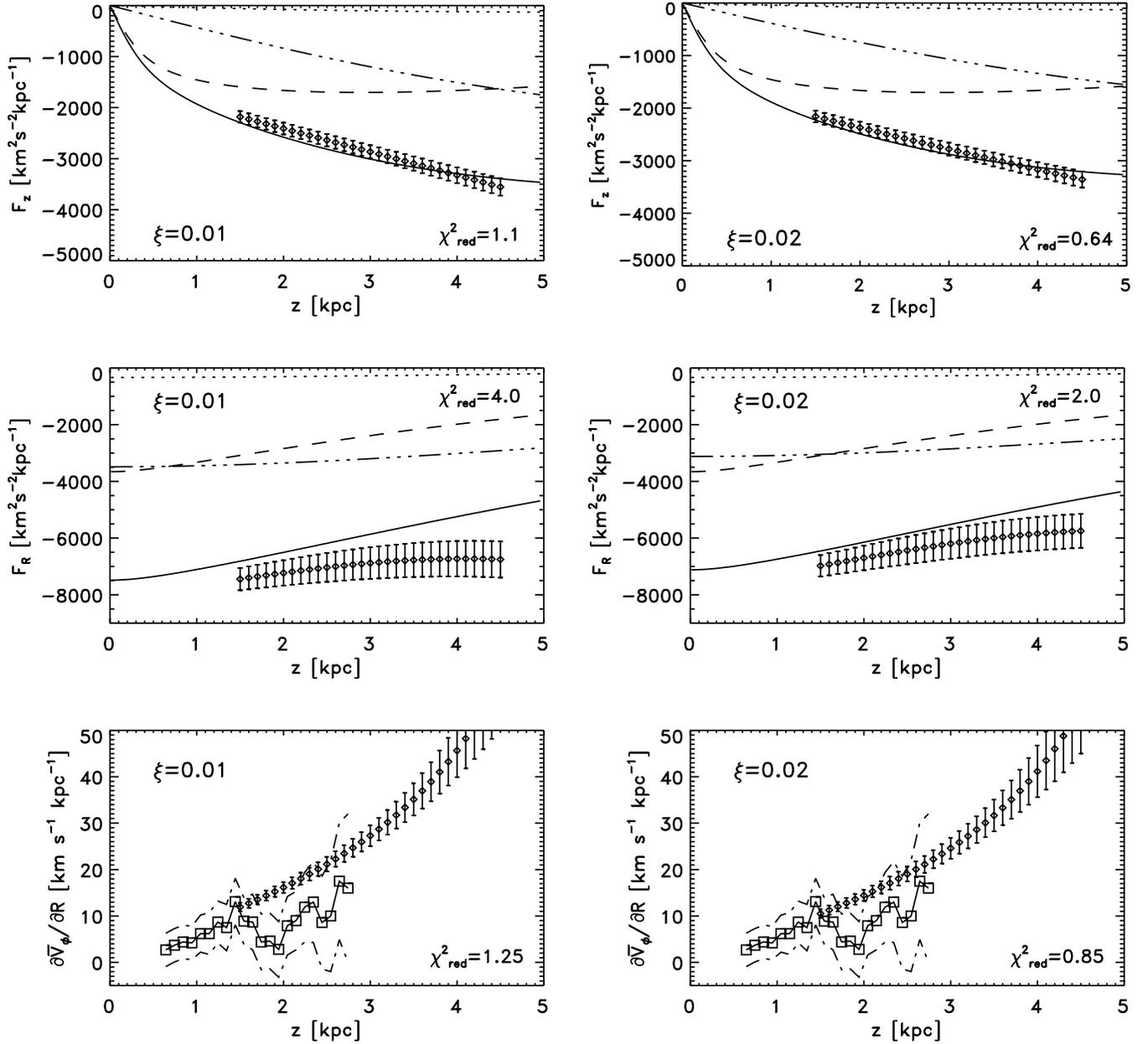}
\caption{Vertical profiles of $F_{z}$, $F_{R}$ and 
$\partial\bar{V}_{\phi}/\partial R$.
In the left column, the solid lines
indicate the profiles for a mass model with $V_{h}=167$ km s$^{-1}$
and $\beta_{h}=0.35$. In the right column, solid curves are 
for a mass model with $V_{h}=158$ km s$^{-1}$ and $\beta_{h}=0.32$.
The contribution of the disk (dashed line),
bulge (dotted line) and halo (triple dotted-dashed line) are also 
shown. 
The dots with error bars indicate the dynamical measurements
for $\xi=0.01$ (left column) and $\xi=0.02$ (right column). 
The dynamical estimates for $F_{z}$, $F_{R}$ and 
$\partial \bar{V}_{\phi}/\partial R$ were calculated
using Equations (\ref{eq:fzest})-(\ref{eq:dVc_dR}). 
In Equation (\ref{eq:dVc_dR}) we used the values of $\Gamma$ of the
corresponding mass model described above.
In the bottom panels, the squares represent the observed value and
the dash-dot lines the $1\sigma$ confidence interval.
For the tracer thick disk population, we have used $h_{R}=3.6$ kpc and
$h_{z}=0.68$ kpc. The $\chi^{2}_{\rm red}$-value is given in each panel. }
\label{fig:best_fits_xi}
\end{figure*}

\begin{figure*} %
\epsscale{0.8}
\plotone{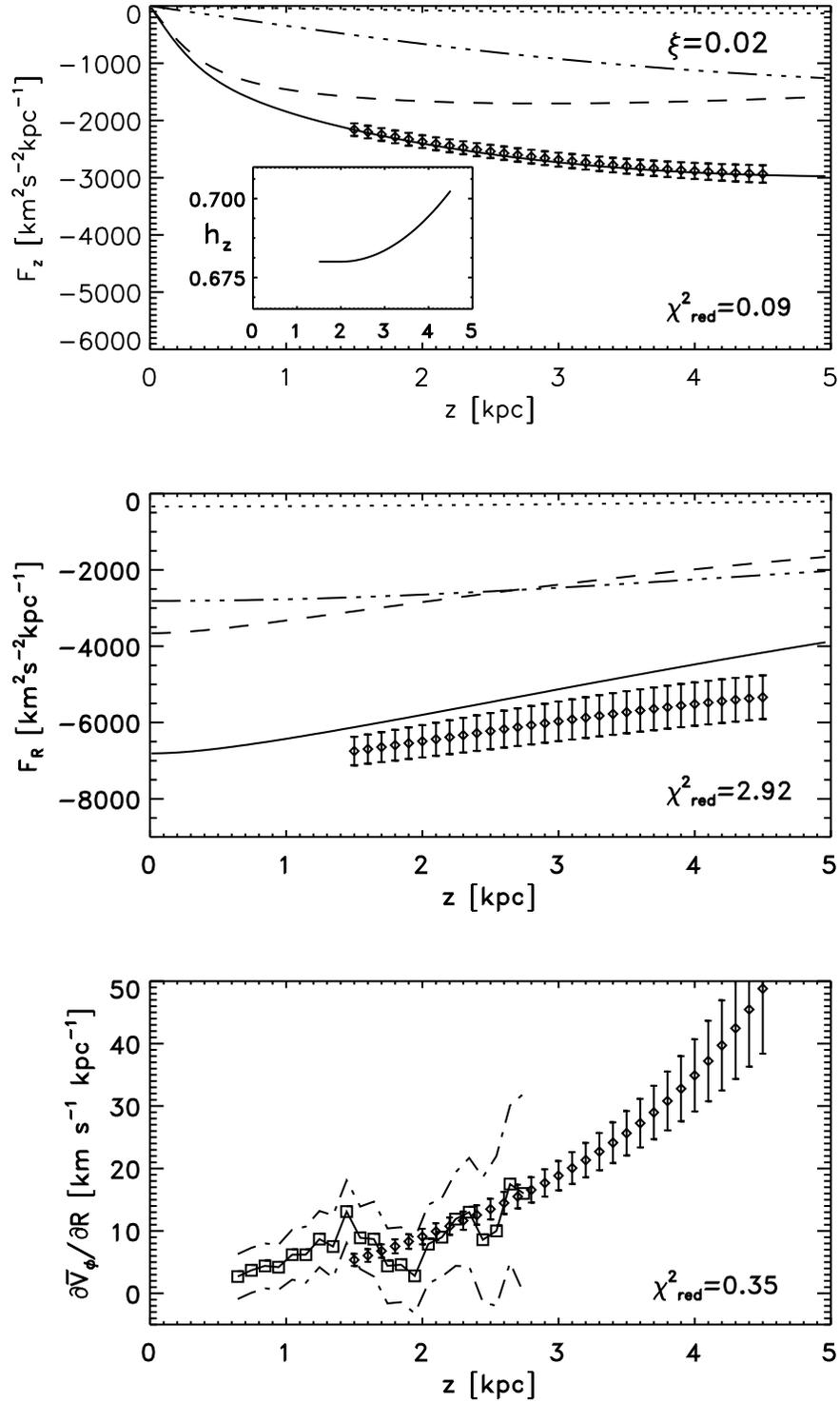}
\caption{Same as Figure \ref{fig:best_fits_xi} but for a mass model
with $V_{h,\odot}=150$ km s$^{-1}$ and $\beta_{h}=0$ (implying that
$V_{c,0}=234$ km s$^{-1}$, $\Gamma_{0}=-2400$ km$^{2}$s$^{-2}$kpc$^{-1}$
and $\rho_{0}=0.0064M_{\odot}$pc$^{-3}$). The scaleheight
depends slightly on $z$, as it is shown in the inset plot at the top panel.}
\label{fig:best_fits_2simultaneous}
\end{figure*}

\begin{figure*} %
\epsscale{0.8}
\plotone{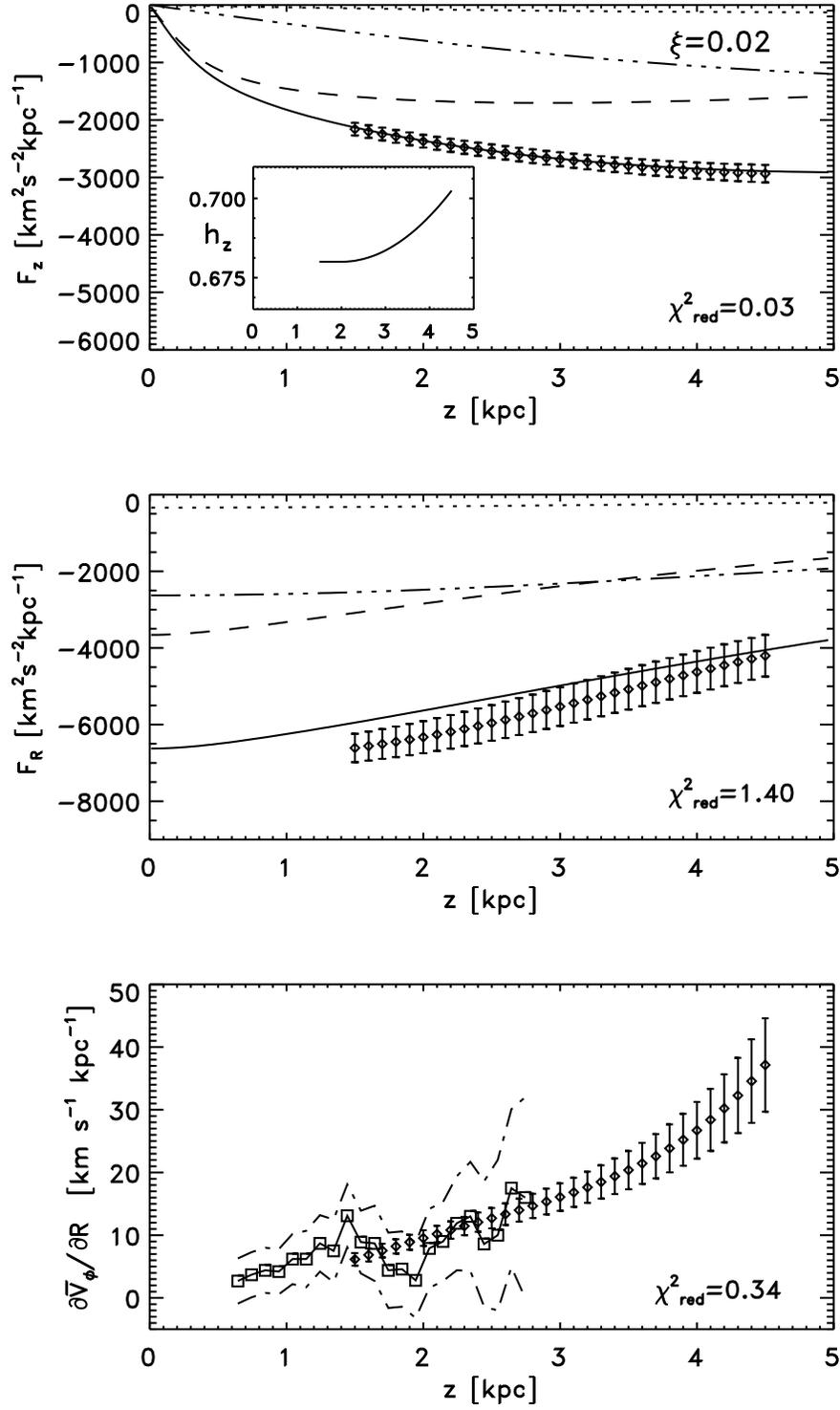}
\caption{
Same as Figure \ref{fig:best_fits_xi} but now for a thick disk with 
a scalelength $h_{\sigma_{R}}$ that increases with $z$, from $h_{\sigma_{R}}=4$ kpc
at $z=1.5$ kpc to $h_{\sigma_{R}}=6.5$ kpc at $z=4$ kpc.
The solid lines represent the profiles  
for a mass model with $V_{h,\odot}=145$ km s$^{-1}$ and $\beta_{h}=0.05$,
resulting in 
$V_{c,0}=231$ km s$^{-1}$, $\Gamma_{0}=-2100$ km$^{2}$s$^{-2}$kpc$^{-1}$
and $\rho_{0}=0.0066M_{\odot}$pc$^{-3}$.}
\label{fig:best_fits_hRz}
\end{figure*}

Knowing $V_{h,\odot}$ and $\beta_{h}$, we may estimate the mean density
of dark matter between $r_{1}$ and $r_{2}$ as
\begin{equation}
\bar{\rho}_{\rm dm}= \frac{3}{4\pi}\frac{M_{h}(r_{2})-M_{h}(r_{1})}{r_{2}^{3}-r_{1}^{3}},
\end{equation}
with $M_{h}(r)$ the DM mass interior to $r$:
\begin{equation}
M_{h}(r)=\frac{rV_{h}^{2}(r)}{G}=\left(\frac{r}{R_{\odot}}\right)^{2\beta_{h}}\frac{rV_{h,\odot}^{2}}{G}.
\end{equation}
The DM density can be estimated as:
\begin{equation}
\rho(r)=\frac{1}{4\pi G r^{2}}\frac{dM_{h}}{dr}=(1+2\beta_{h})\frac{V_{h}^{2}(r)}
{4\pi G r^{2}}.
\end{equation}
In particular, the local DM density is:
\begin{equation}
\rho_{0}=(1+2\beta_{h})\frac{V_{h,\odot}^{2}}
{4\pi G R_{\odot}^{2}}.
\end{equation}

\subsection{Results}
To estimate $F_{z}^{\rm est}$, $F_{R}^{\rm est}$ and $\Gamma^{\rm est}$,
we need to set the geometrical parameters of the thick disk.
To facilitate comparison with Bovy \& Tremaine (2012) and 
Moni Bidin et al.~(2015),
we start by considering a model with $h_{R}=3.6$ kpc, which
corresponds to the scalelength derived by Juri\'c et al.~(2008). They
estimated an error in $h_{R}$ of $20\%$. 
As far as we know, the only measurements of $h_{\sigma_{R}}$ and
$h_{\sigma_{\phi}}$ were provided by Lewis \& Freeman (1989).
They estimated $h_{\sigma_{R}}=4.37\pm 0.32$ kpc and
$h_{\sigma_{\phi}}=3.36\pm 0.62$ kpc (for $R_{\odot}=8.5$ kpc
and $V_{c,0}=220$ km s$^{-1}$).
For $\sigma^{2}_{z}$, Bovy et al.~(2012a) 
found a scalelength of $3.5$ kpc, whereas Hattori \& Gilmore (2015)
derived a value of $4.1\pm 1$ kpc.
Following Bovy \& Rix (2013), we
set $h_{\sigma_{R}}=h_{\sigma_{\phi}}=4$ kpc.
For $h_{\sigma_{Rz}}$ we adopt a value of $3.5$ kpc,
similar to the scalelength observed for $\sigma_{z}$ by Bovy et al.~(2012a).

Values for the flare parameter of the thick disk, $\xi$, between $0.01$ and 
$0.02$ were found by Mateu et al.~(2011) and by 
L\'opez-Corredoira \& Molg\'o (2014). Therefore, we perform the analysis
assuming that the flaring is linear and consider two values $\xi=0.01$
and $\xi=0.02$.

Once $h_{R}$, $h_{\sigma_{R}}$, $h_{\sigma_{\phi}}$, $h_{\sigma_{Rz}}$
and $\xi$ are fixed, we only need to set $h_{z}$ to measure
$F_{z}^{\rm est}$, $F_{R}^{\rm est}$ and 
$\Gamma^{\rm est}$, using 
Eqs.~(\ref{eq:fzest})-(\ref{eq:dVc_dR}).
Then, we can adjust $V_{h,\odot}$ and $\beta_{h}$ of our mass model
to match {\it simultaneously} $F_{z}^{\rm est}$, $F_{R}^{\rm est}$
and $\Gamma^{\rm est}$. The local circular velocity
corresponding to the halo,
$V_{h,\odot}$, is not permitted to be larger than $174$ km s$^{-1}$
to guarantee that $V_{c,0}<250$ km s$^{-1}$.
We comment that the local surface density is uniquely determined
by $F_{z}$ and $\Gamma$ (see Equation \ref{eq:init}).
Indeed, two models having the same $F_{z}$ and $\Gamma$ should have
the same surface density even if they have different $F_{R}$.
Still, it is useful to fit also $F_{R}^{\rm est}$ to reduce
uncertainties in the parameters.

We find that the quality of the fits depends on the selected value
for $h_{z}$;
the best fits are obtained when $h_{z}$ lies between $0.64$ kpc and
$0.7$ kpc. These values of $h_{z}$ are consistent with the scale height derived
in Bilir et al.~(2008), Polido et al.~(2013), Jia et al.~(2014)
and L\'opez-Corredoira et al.~(2015). Other authors measure scaleheights
up to $1.0$ kpc (for a compilation, se Table 1 in Jia et al.~2014). 

Figure \ref{fig:best_fits_xi} shows our best fits for $h_{z}=0.68$ kpc. For
$\sigma_{Rz}^{2}$, we have used the fit given in Equation 
(\ref{eq:sigmaRz2}). Instead of $\Gamma$, we show
$\partial \bar{V}_{\phi}/\partial R$ as derived using Equation
(\ref{eq:dVc_dR}) and compare it with the observed measurements. 
The reduced chi-square, $\chi_{\rm red}^{2}$, are quoted at a corner in
each panel. 

For $h_{z}=0.68$ kpc and $\xi=0.01$ (and using our fiducial values
of the geometrical parameters), we find $V_{h,\odot}=167\pm 2$ km s$^{-1}$
and $\beta_{h}=0.35\pm 0.10$ (thus $\rho_{0}=0.0135\pm 0.002M_{\odot}$
pc$^{-3}$). For the same $h_{z}$ but
for $\xi=0.02$, we get $V_{h,\odot}=158\pm 2$ km s$^{-1}$
and $\beta_{h}=0.32\pm 0.10$ (hence $\rho_{0}=0.012\pm 0.002M_{\odot}$
pc$^{-3}$).  
These models have local circular velocities at
the midplane of $\sim 240$ km s$^{-1}$, and small values for $\Gamma_{0}$
(between $-500$ km$^{2}$s$^{-2}$kpc$^{-1}$ and 
$100$ km$^{2}$s$^{-2}$kpc$^{-1}$), which are consistent with the values 
derived in \S \ref{sec:representative} using the Oort constants.

The goodness of the fits to $F_{z}^{\rm est}$
is statistically similar for $\xi=0.01$ and $\xi=0.02$.
The curves $F_{z}^{\rm est}$ vs $z$ are linear in both cases, 
while the vertical profile $F_{z}$ of the best-fitting mass 
models present some curvature. 
If the thick disk is not strictly exponential along $z$ (e.g.,
\S \ref{sec:dynamicaltests} and Carollo et al.~2010), but $h_{z}$ 
increases slightly from $z=1.5$ kpc to $z=4.5$ kpc,
then $F_{z}^{\rm est}$ vs $z$ is not longer linear but flattens.
A small deviation from a strict vertical exponential profile
provides a good fit to the $z$-component of the force (see below).

In \S \ref{sec:bt_vs_cMB}, we showed that, when $\xi=0$, the shape of 
$F_{R}^{\rm est}$ vs $z$ is unphysical. However, for $\xi=0.02$, the shape 
of $F_{R}^{\rm est}$ vs $z$ appears reasonable (see Figure \ref{fig:best_fits_xi}). 
Still, the model underestimates $|F_{R}|$ over the errors.
It is worthwhile noticing that the slope of $F_{R}$ in the mass 
models is essentially given by the baryonic disk because the contribution
of the halo is almost flat.
If we adopt a larger value for $V_{h,\odot}$, $|F_{R}|$ of the mass
model increases but, since $F_{R}^{\rm est}$ depends implicitly
on $V_{c,0}$ through the term $\bar{V}_{\phi}^{2}/R$ (see Equations
\ref{eq:fRest} and \ref{eq:Vphi}), a larger $V_{h,\odot}$ also implies a higher 
$|F_{R}^{\rm est}|$. Thus, the convergence $F_{R}\rightarrow F_{R}^{\rm est}$
by changing $V_{h,\odot}$ is very slow. In fact,
we need $V_{h,\odot}\simeq 190$ km s$^{-1}$ and, thereby
$V_{c,0}\simeq 260$ km s$^{-1}$ to account for $F_{R}^{\rm est}$.  
Since this value for $V_{c,0}$ is not very plausible,
the lack of a mass model able to fit $F_{R}^{\rm est}$ 
is telling us that either the selected geometrical
parameters are not correct, the flaring of the disk is higher than
we assumed or, more likely, the kinematical data are biased 
(or a combination).

The vertical profile for $\partial \bar{V}_{\phi}/\partial R$, derived
using Equation (\ref{eq:dVc_dR}), 
is compared with the observed values of this quantity
by Moni Bidin et al.~(2015). We see that although the values
of $\chi^{2}_{\rm red}$ are close to $1$, the estimated values
of $\partial \bar{V}_{\phi}/\partial R$ systematically lie over the
observed values, which is not very satisfactory.

Satisfactory fits to $\partial \bar{V}_{\phi}/\partial R$, can be achieved 
using a somewhat lower value for $\beta_{h}$. 
Figure \ref{fig:best_fits_2simultaneous} shows a mass model 
which was tailored to fit both $F_{z}$ and $\partial \bar{V}_{\phi}/\partial R$.
The mass model has $\rho_{0}=0.0064M_{\odot}$pc$^{-3}$.
In this fit, we have used a $z$-dependent vertical scale height
to improve the $F_{z}$ fit. We see that the $F_{R}$-fit is slightly
worsened and, more importantly, the value of $\Gamma_{0}$ in this mass
model is $-2400$ km$^{2}$s$^{-2}$kpc$^{-1}$, which is just outside
the range derived using measurements of the Oort constants in 
\S \ref{sec:representative}.

If the assumption that $h_{\sigma_{ij}}$ are constant with $z$ are relaxed,
the $F_{R}$-fit can be improved. Our mock thick disk described
in \S \ref{sec:sims} suggests that all $h_{\sigma_{ij}}$ 
increase with $z$, particularly $h_{\sigma_{R}}$. For illustration, 
Figure \ref{fig:best_fits_hRz} shows
the case where $h_{\sigma_{R}}$ increases from $4$ kpc at $z=1.5$ kpc
to $6.5$ kpc at $z=4$ kpc in the following manner:
\begin{equation}
h_{\sigma_{R}}[{\rm kpc}]=4+0.4(z-1.5)^{2}.
\label{eq:hsigmaRvsz}
\end{equation}
The halo parameters for this model are $V_{h,\odot}=145\pm 3$ km s$^{-1}$ and 
$\beta_{h}=0.05\pm 0.1$. In this model, the local DM density, 
$\rho_{0}$, is $0.0066\pm 0.0015 M_{\odot}$ pc$^{-3}$.
The value for $\chi_{\rm red}^{2}$ of the $F_{R}$-fit is a bit less 
($\simeq 1.4$), but not fully satisfactory. 

In order to explore the sensitivity of the fits to the adopted values
of the geometrical parameters, we recalculated the best-fitting halo parameters
using $h_{\sigma_{R}}$ as given in Equation (\ref{eq:hsigmaRvsz}), but 
$h_{\sigma_{\phi}}=h_{\sigma_{Rz}}=4.15$ kpc, as derived by 
Hattori \& Gilmore (2015) for $h_{\sigma_{z}}$. The best fitting 
parameters are $V_{h,\odot}=142\pm 3$ km s$^{-1}$
and $\beta_{h}=0.13\pm 0.1$, implying $\rho_{0}=0.0073\pm 0.0015 M_{\odot}$ pc$^{-3}$
The value of $\Gamma_{0}$ is $-1730$ 
km$^{2}$s$^{-2}$kpc$^{-1}$, which is reasonable. 

The quoted errors in $\rho_{0}$ do not include
uncertainties in the level of flattening of the dark halo, in 
the gas and stellar surface density, in the value of $R_{\odot}$, or 
associated with the assumption that the tracers are exponentially 
distributed in radius and height (see Hessman 2015 and McKee et al.~2015,
for a recent discussion of these issues).

\subsection{Discussion}
In the previous subsection we found that acceptable fits to 
$F_{z}^{\rm est}$, $F_{R}^{\rm est}$ and $\Gamma^{\rm est}$ are obtained
for $h_{z}\simeq 0.68$ kpc and $h_{R}\gtrsim 3.6$ kpc. These models
have local circular velocities of the halo, $V_{h,\odot}$, 
in the range $140$-$155$ km s$^{-1}$ and the $\beta_{h}$ parameter
between $0$ and $0.15$. In all these 
models, $\rho_{0}\gtrsim 0.0064\pm 0.0015 M_{\odot}$ pc$^{-3}$.
Fits of the same quality can be achieved if both $h_{z}$ and $h_{R}$ 
are reduced simultaneously; for instance, taking $h_{z}=0.64$ kpc 
and $h_{R}=3.4$ kpc, we can find fits of the same quality, although a 
slightly larger $\rho_{0}$ is required. 

While these values for $h_{z}$ and $h_{R}$ seem 
reasonable [e.g., de Jong et al.~(2010) found 
$h_{R}=4.1\pm 0.4$ kpc and $h_{z}=0.75\pm 0.07$ kpc 
and Bilir et al.~(2008) derived a height of $0.55-0.72$ kpc 
assuming $h_{R}=3.5$ kpc; see also Polido et al.~(2013)
for similar values], 
recent studies with a larger cover in altitude and longitude suggest 
a shorter scalelength, $h_{R}\sim 2.3$ kpc, for the thick disk 
(Robin et al.~2014 and references therein).
Therefore, it is relevant to consider models with a smaller value for
$h_{R}$.

\begin{figure} %
\epsscale{1.26}
\plotone{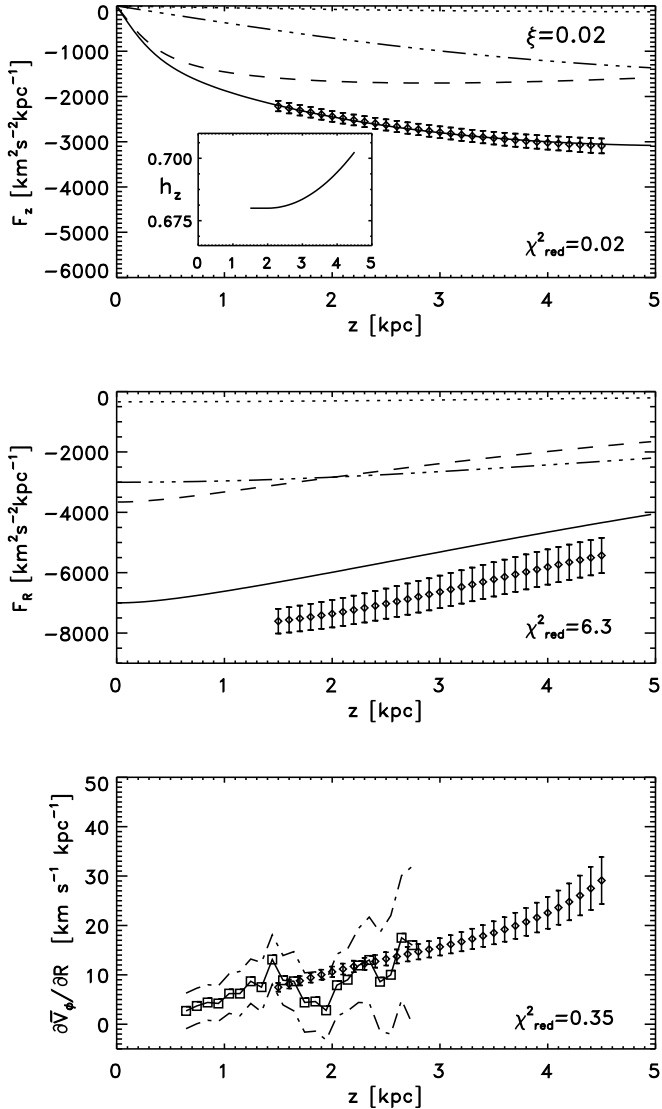}
\caption{
Vertical profiles of $F_{z}^{\rm est}$, $F_{R}^{\rm est}$ and 
$\Gamma^{\rm est}$ for the following
geometrical parameters: $h_{R}=2.5$ kpc, $h_{\sigma_{R}}$ as
given in Equation (\ref{eq:hsigmaRvsz}), 
$h_{\sigma_{\phi}}=h_{\sigma_{Rz}}=4.15$ kpc. 
The mass model has $V_{h,\odot}=155$ km s$^{-1}$ and $\beta_{h}=0.05$.
Thus, $V_{c,0}=237$ km s$^{-1}$, $\Gamma_{0}=-2080$ km$^{2}$s$^{-2}$kpc$^{-1}$
and $\rho_{0}=0.0076M_{\odot}$pc$^{-3}$.
}
\label{fig:hR25hz9}
\end{figure}

Figure \ref{fig:hR25hz9} shows a case with $h_{R}=2.5$ kpc, $h_{\sigma_{R}}$
following Equation (\ref{eq:hsigmaRvsz}) and 
$h_{\sigma_{\phi}}=h_{\sigma_{Rz}}=4.15$ kpc. 
The flaring parameter was also taken $\xi=0.02$.
The reduction in $h_{R}$ leads to higher values for both $|F_{z}^{\rm est}|$
and $|F_{R}^{\rm est}|$, thus $\rho_{0}$ increases.
If the fits to $F_{z}^{\rm est}$ and $\partial\bar{V}_{\phi}/\partial R$
are forced to be as good as in the previous cases, the $F_{R}$
fit becomes poorer.  In this case, the best-fitting halo parameters 
are $V_{h,\odot}=155$ km s$^{-1}$ and
$\beta_{h}=0.05$ ($\rho_{0}=0.0076M_{\odot}$pc$^{-3}$).

As already said in the previous Subsection, the local surface
density is uniquely determined by $F_{z}$ and $\Gamma$.
In order to derive the local surface density, we only need to know
$F_{z}$ and $\Gamma$. Thus, a bad fit to $F_{R}$ is not critical 
regarding the estimate the local volume density because the Poisson
equation does not depend on $F_{R}$; if the kinematical data is reliable,
good fits to $F_{z}$ and $\partial\bar{V}_{\phi}/\partial R$ would be enough 
to infer the local volume density of mass. Taken at face value, a mismatch
between $F_{R}^{\rm est}$ and $F_{R}$ in the model could be
indicative that the halo is not spherical interior to $R_{\odot}$.
Still, it turns out that the value of $V_{c,0}$ required to
account for $F_{R}^{\rm est}$, as derived for $h_{R}=2.5$ kpc, is 
excessively large, even if the halo is prolate.
This may indicate the kinematical data is contaminated by systematic
uncertainties.

In conclusion, for $h_{R}\sim 2.5$ kpc, 
$F_{R}^{\rm est}$ cannot be fitted satisfactorily, 
unless $V_{c,0}$ is unusually large. Thus, we suggest that much
of the offset between $F_{R}^{\rm est}$ and what models predict is
caused by systematics in the data. If we fit simultaneously $F_{z}^{\rm est}$
and $\Gamma^{\rm est}$, and ignore $F_{R}^{\rm est}$ (because it does not
participate in the Poisson equation) then we obtain 
$\rho_{0}=0.0076\pm 0.002M_{\odot}$pc$^{-3}$ for $h_{z}\simeq 0.68$ kpc. 
As already said, a reduction of $h_{R}$ results in a higher value
for $\rho_{0}$ because $|F_{z}^{\rm est}|$ increases, whereas 
$\Gamma^{\rm est}$ is independent of $h_{R}$.
Moni Bidin et al.~(2012b) found that large values for $h_{R}$,
typically $h_{R}\sim 4.7$ kpc, were required to recover the standard 
values for $\rho_{0}$. This result is a consequence of their different 
assumptions; mainly their hypothesis that $h_{R}=h_{\sigma}$.

So far, we have assumed that the baryonic disk can be described
by an exponential disk with a mass-weighted scale length, $R_{d}$,
of $2.1$ kpc. If $R_{d}$ is larger, the baryonic mass in the disk
turns to be smaller and hence more DM is required to account for
the dynamics. Typically, $\rho_{0}$ increases by $20\%$,
if we adopt $R_{d}=3$ kpc instead of $2.1$ kpc.

\section{Summary and conclusions}
\label{sec:conclusions}

Having knowledge of the volume density $\nu(R_{\odot},z)$ and vertical
velocity for a vast number of some equilibrium tracer stars, the 
dynamical surface density can be measured at heights $|z|<1.5$ kpc above
the Galactic midplane, using the one-dimensional approximation, 
i.e.~$S_{F_{R}}\simeq 0$ and neglecting the $\sigma_{Rz}^{2}$-tilt term
in the Jeans equation (e.g., Kuijken \& Gilmore 1989; Read 2014).
At higher Galactic heights, the one-dimensional approximation 
may introduce bias (Siebert et al.~2008; Smith et al.~2009; 
Garbari et al.~2012; McKee et al.~2015). Bovy \&
Tremaine (2012) claimed that the assumption $S_{F_{R}}\simeq 0$ provides a 
robust lower limit to the surface density at $|z|>1.5$ kpc, as far as 
$\partial V_{c}/\partial R\simeq 0$ in the midplane.  
Using the sample of thick disk stars reported in Moni Bidin et al.~(2012a), 
Bovy \& Tremaine (2012) found that the mean halo density between $z=1$ and 
$z=4$ kpc is $0.0095\pm 0.0015 M_{\odot}$pc$^{-3}$ for $h_{R}=2$ kpc and 
$h_{z}=0.7$ kpc or $0.007\pm 0.001 M_{\odot}$pc$^{-3}$ for $h_{R}=3.8$ kpc 
and $h_{z}=0.9$ kpc.

Moni Bidin et al.~(2015) have challanged the $S_{F_{R}}\geq 0$ assumption
of Bovy \& Tremaine (2012) on the basis that it is not generally true that it holds
for any Galactic potential. Moni Bidin et al.~(2015) argue that this assumption is 
implicitly constraining the mass distribution and, thus, it is not adequate to derive
$\rho_{0}$. In order to release this hypothesis, Moni Bidin et al.~(2015)
moved beyond the one-dimensional approximation
and, instead of assuming $S_{F_{R}}\simeq 0$ (or, equivalently 
$\Gamma\simeq 0$), they used a dynamical measure
of $\Gamma$ as a function of $z$, 
which was derived from the kinematics of the same tracer population.  
The computation of $\Gamma$ requires knowledge of
$\bar{V}_{\phi}$, which implicitly depends on the local circular 
velocity at the midplane, and $\partial \bar{V}_{\phi}/\partial R$, 
which is rather uncertain.  Using this three-dimensional formalism, 
Moni Bidin et al.~(2015) found that 
the density of DM at $|z|>1.5$ kpc is $0\pm 0.002 M_{\odot}$
pc$^{-3}$ for $h_{R}=2$ kpc and $h_{z}=0.7$ kpc 
and $0.002\pm 0.003 M_{\odot}$ pc$^{-3}$ for $h_{R}=3.6$ kpc and 
$h_{z}=0.9$ kpc. These results are at variance with those derived by Bovy 
\& Tremaine (2012).
In order to fully understand this discrepancy, we investigated
the robustness of both approaches in detail.

We have shown that the cMB estimator (after correcting it by a missed
term in Moni Bidin et al.~2012b, 2015) is more sensitive to uncertainties
in the geometrical parameters of the thick disk, particularly
on $h_{\sigma}$, than the BT estimator. Therefore, accurate determinations
of the dynamical surface density require good measures of the
radial scalelengths of each of the components of the velocity 
dispersion tensor. 

Using the Jeans equations and the current data of the kinematics of thick
disk stars, we derived the vertical profiles of $F_{z}$, $F_{R}$ and $\Gamma$.
Assuming $V_{c,0}=215$ km s$^{-1}$ and $h_{\sigma}=3.5$ kpc,
we found that the kinematical estimates of $F_{R}$ and $\Gamma$ are biased towards
large negative values. Possible causes are:
poor model assumptions (e.g., neglecting the flare of the thick disk),
uncertainties in the adopted values for $V_{c,0}$, $\bar{V}_{\phi}$ or 
$\partial \bar{V}_{\phi}/\partial R$,
systematic overestimations of $\sigma_{R}$, $\sigma_{\phi}$, or $\sigma_{Rz}$,
or underestimations of the radial scalelengths of the components
of the velocity dispersion tensor.

We have built simple but representative Galactic
models consisting of the bulge, the baryonic disk and a DM component,
having a local DM density as those 
inferred by the BT estimator and the cMB estimator (using
$h_{\sigma}=3.5$ kpc and $V_{c,0}=215$ km s$^{-1}$. 
We have compared $F_{z}$, $F_{R}$ and $\Gamma$ in these Galactic
mass models with their corresponding estimates using the
currently available kinematical data of the thick disk and found significant
discrepancies between them.
This implies that either the mass models do not describe
correctly the underlying Galactic potential, or systematic
errors and uncertain assumptions are affecting the dynamical estimates
of $F_{z}$, $F_{R}$ and $\Gamma$. We also stress the result
that mass distributions with different local surface density
may exhibit similar values of both $F_{z}$ and $F_{R}$. To
discriminate between these models, precise measurements of $\Gamma_{0}$
are required. Since the currently available kinematical data of thick disk stars
are not good enough to provide a precise value of $\Gamma_{0}$, 
we conclude that these data are not sufficient to constrain 
$\Sigma (z)$ without making further assumptions.

In order to examine the role and magnitude of the different systematics, we applied
the BT and cMB estimators to estimate the surface density of a mock thick disk
of tracer stars embedded in a fixed Milky Way-like gravitational potential, and compared
with the exact input value.
To do so, we prepared a thick disk population as close as we could to the assumptions
behind the three-dimensional approach, namely,
a thick disk following a rather homogeneous, doubly exponential distribution
of stars in radius and height. 
However, our simulated thick disk has $h_{\sigma_{ij}}$ that depend
strongly on $z$, whereas the cMB apprach assumes that they are constant
with $z$.
Our analysis suggests that, even if $V_{c,0}$,
$\bar{V}_{\phi}$, $\partial\bar{V}_{\phi}/\partial R$ and
all the mean geometrical parameters of the thick disk were accurately known,
inaccuracies underpredict $\Gamma$ by a factor of 
$3$ at $z=3.5$ kpc above the disk. Still, the cMB formula, applied to our 
mock data, was able to recover the input surface density within
the uncertainties.

Under the assumption that the DM halo is spherically symmetric interior
to the radius $\sim 1.15R_{\odot}$, 
we have explored what configuration of parameters is more
compatible with the current kinematic data of the thick disk.
We have used a parametric mass model, consisting of the bulge,
the baryonic disk and a dark halo. 
We have fit $F_{z}^{\rm est}$, $F_{R}^{\rm est}$ and $\Gamma^{\rm est}$
with two free parameters; the contribution of the dark halo to the
local circular velocity $V_{h,\odot}$ and the power-law exponent
$\beta_{h}$. Acceptable fits are obtained if the thick disk is
slightly flared and for the following geometrical parameters:
$h_{z}\sim 0.7$ kpc, $h_{R}\gtrsim 3.5$ kpc, $h_{R}\gtrsim 4$ kpc,
$h_{\sigma_{\phi}}\gtrsim 4$ kpc, and for a dark halo with
$\rho_{0}\gtrsim 0.0064M_{\odot}$pc$^{-3}$.
For smaller values of $h_{R}$, the estimated value for $\rho_{0}$
increases, but the fit to $F_{R}$ worsens.

The currently available kinematical data of the thick disk stars 
up to $4$ kpc
above the Galactic midplane, may be also compatible with a flatten 
disk-like distribution of DM, having a DM density at $|z|>1.5$ kpc as low
as that derived in Moni Bidin et al.~(2015). However, this model possibly has
too large negative values of $\Gamma_{0}$. 

\acknowledgements
We thank the referee, Dr.~C.~Moni Bidin, for a very detailed 
report and constructive comments, which helped to improve the 
quality of the paper significantly.
This work has been partly supported by CONACyT project 165584.

\appendix
\section{A. On the second-order derivatives in the Jeans equations}
\label{sec:second_order}
In \S \ref{sec:Jeans_Poisson}, we used the Jeans equations to obtain
the formulae for $F_{z}$, 
$F_{R}$ and $\Gamma$ as a function of the velocity dispersions and
scalelengths of a tracer population. We considered that
the tracer population is a flaring stellar disk, and assumed
that the velocity dispersion components have exponential profiles 
along the $R$-direction.
It is useful to derive $F_{z}$, $F_{R}$ and $\Gamma$ in a general form.
We define the generalized geometrical factors, which depend on $(R,z)$,
as:
\begin{equation}
\tilde{h}_{z}^{-1}\equiv -\frac{1}{\nu}\frac{\partial \nu}{\partial z},
\hskip 2.0cm
\tilde{h}_{R}^{-1}\equiv -\frac{1}{\nu}\frac{\partial \nu}{\partial R},
\hskip 2.0cm
H_{R}^{-2}\equiv \frac{1}{\nu}\frac{\partial^{2} \nu}{\partial R^{2}},
\end{equation}
and
\begin{equation}
\tilde{h}_{\sigma_{ij}}^{-1}\equiv -\frac{1}{\sigma_{ij}^{2}}
\frac{\partial \sigma_{ij}^{2}}{\partial R},
\hskip 2.0cm
H_{\sigma_{ij}}^{-2}\equiv \frac{1}{\sigma_{ij}^{2}}
\frac{\partial^{2} \sigma_{ij}^{2}}{\partial R^{2}}.
\end{equation}
Equations (\ref{eq:fz}) and (\ref{eq:fR}) can be written as 
\begin{equation}
F_{z}=\frac{\partial \sigma_{z}^{2}}{\partial z}-\frac{\sigma_{z}^{2}}{\tilde{h}_{z}}
+\tilde{k}_{0}\sigma_{Rz}^{2},
\end{equation}
with $\tilde{k}_{0}\equiv R^{-1}-\tilde{h}_{R}^{-1}-
\tilde{h}_{\sigma_{Rz}}^{-1}$ and
\begin{equation}
F_{R}=\tilde{k}_{0}'\sigma_{R}^{2}-\frac{1}{R}(\sigma_{\phi}^{2}+\bar{V}_{\phi}^{2})
+\frac{\partial \sigma_{Rz}^{2}}{\partial z}-\frac{\sigma_{Rz}^{2}}{\tilde{h}_{z}},
\end{equation}
with $\tilde{k}_{0}'\equiv R^{-1}-\tilde{h}_{R}^{-1}-\tilde{h}_{\sigma_{R}}^{-1}$. Finally, $\Gamma$ is given by:
\begin{equation}
-\Gamma= K_{1}\sigma_{R}^{2}+
\frac{\sigma_{\phi}^{2}}{\tilde{h}_{\sigma_{\phi}}}\pm K_{2}R 
\frac{\sigma_{Rz}^{2}}{\tilde{h}_{z}}
+\frac{\partial \sigma_{Rz}^{2}}{\partial z}
+R\frac{\partial^2 \sigma_{Rz}^{2}}{\partial z\partial R}
-\frac{\partial{\bar{V}_{\phi}^{2}}}{\partial R},
\end{equation}
where 
\begin{equation}
K_{1}=\left(\frac{1}{H_{R}^{2}}-\frac{1}{\tilde{h}_{R}^{2}}+
\frac{1}{\tilde{h}_{R}\tilde{h}_{\sigma_{R}}}+
\frac{1}{H_{\sigma_{R}}^{2}}\right)R
-\frac{1}{\tilde{h}_{R}}-\frac{2}{\tilde{h}_{\sigma_{R}}},
\end{equation}
and
\begin{equation}
K_{2}=\frac{1}{\tilde{h}_{\sigma_{Rz}}}-\frac{1}{R}-\frac{1}{\tilde{h}_{R}}
+\frac{\tilde{h}_{z}}{\nu}\frac{\partial^{2}\nu}{\partial z\partial R}.
\end{equation}

The geometrical parameters $\tilde{h}_{z}$, $\tilde{h}_{R}$ and 
$\tilde{h}_{\sigma_{ij}}$ are essentially local derivatives of $\nu$ and 
$\sigma_{ij}^{2}$, which can be measured in the simulations
at a generic point $(R,z)$.
However, in \S \ref{sec:dynamicaltests}, 
we computed $\Gamma$ assuming that $H_{R}=\tilde{h}_{R}$ and 
$H_{\sigma_{R}}=\tilde{h}_{\sigma_{R}}$, which is only correct
when $\nu$ and $\sigma_{R}^{2}$ have strict exponential profiles 
with $R$. In particular, if $\nu$ and $\sigma_{\sigma_{R}}^{2}$ have
exponential profiles in the $R$-direction, then $H_{R}=\tilde{h}_{R}$,
$H_{\sigma_{R}}=\tilde{h}_{\sigma_{R}}$, and hence
\begin{equation}
K_{1}=\left(\frac{1}{\tilde{h}_{R}\tilde{h}_{\sigma_{R}}}+
\frac{1}{\tilde{h}_{\sigma_{R}}^{2}}\right)R
-\frac{1}{\tilde{h}_{R}}-\frac{2}{\tilde{h}_{\sigma_{R}}},
\end{equation}
and
\begin{equation}
K_{2}=\frac{1}{\tilde{h}_{\sigma_{Rz}}}-\frac{1}{R}.
\end{equation}

If $\nu$ and $\sigma_{R}^{2}$ are not exponential but
follow a power-law along $R$, such as 
\begin{equation}
\nu=\nu_{\odot}(z)\left(\frac{R_{\odot}}{R}\right)^{l_{\nu}},
\end{equation}
and
\begin{equation}
\sigma_{R}^{2}=\sigma_{R,\odot}^{2}(z)\left(\frac{R_{\odot}}{R}\right)^{l_{\sigma}},
\end{equation}
we get
\begin{equation}
K_{1}= [l_{\nu}
+l_{\sigma}(1+l_{\sigma})]\frac{1}{R}
-\frac{1}{\tilde{h}_{R}}-\frac{2}{\tilde{h}_{\sigma_{R}}},
\end{equation}
and
\begin{equation}
K_{2}=\frac{1}{\tilde{h}_{\sigma_{Rz}}}+\frac{l_{\nu}-1}{R}
-\frac{1}{\tilde{h}_{R}}.
\end{equation}

In the simulations, it is very difficult to measure $H_{R}$ and 
$H_{\sigma_{R}}$, particularly at high $z$. For instance, consider
the $R$-profile of $\sigma_{R}^{2}$.
Between $R=4$ kpc and $R=12$ kpc, $\sigma_{R}^{2}$ vs $R$ in a cut
along $z=3.5$ kpc, can be fitted equally well with an exponential
profile with $\tilde{h}_{\sigma_{R}}=30$ kpc than with a power-law
profile with $l_{\sigma}=0.25$. In the first case $H_{\sigma_{R}}=30$ kpc,
whereas in the second case $H_{\sigma_{R}}=14.5$ kpc.
This uncertainty in the second-order derivatives were not included
in the analysis of \S \ref{sec:dynamicaltests}.

\section{B. Miyamoto-Nagai disk plus a spherical halo}
\label{sec:introproblems}

Consider an axisymmetric distribution of mass consisting of
a Miyamoto \& Nagai (1985) disk and a spherically-symmetric component.
The gravitational potential is given by the sum of the 
contribution of the disk $\Phi_{d}$ plus the contribution of the spherical
component $\Phi_{sph}$. 
We define the corresponding circular velocities of the disk
and spherical component (halo$+$bulge) as
\begin{equation}
V_{c,d}^{2}(R,z)\equiv R\frac{\partial\Phi_{d}}{\partial R},
\end{equation}
and
\begin{equation}
V_{c,sph}^{2}(R,z)\equiv R\frac{\partial\Phi_{sph}}{\partial R}.
\end{equation}
Thus $V_{c}^{2}=V_{c,d}^{2}+V_{c,sph}^{2}$.  We aim to compute how
$\Gamma$ varies with $z$ at a given $R$.

After some simple algebraic manipulations, we find
\begin{equation}
\frac{\partial V_{c,sph}^{2}}{\partial R} = \frac{R}{r^{2}} 
\left(\frac{R^{2}+2z^{2}}{r}\frac{d\Phi_{sph}}{dr} +R^{2}
\frac{d^{2}\Phi_{sph}}{dr^{2}} \right),
\end{equation}
where $r^{2}=R^{2}+z^{2}$.
In addition to the circular velocity $V_{c,sph}$,
it is useful to define the circular velocity of a tilted orbit in the spherical potential, 
\begin{equation}
V_{\rm tilt}^{2}\equiv r \frac{d\Phi_{sph}}{dr}.
\end{equation}
Using this definition, we can rewrite
\begin{equation}
\frac{\partial V_{c,sph}^{2}}{\partial R} = 
\frac{2Rz^{2}}{r^{4}}V_{\rm tilt}^{2}
+\frac{R^{3}}{r^{3}}\frac{dV_{\rm tilt}^{2}}{dr}
\end{equation}
or, equivalently,
\begin{equation}
\frac{\partial V_{c,sph}^{2}}{\partial R} = 
\frac{2R}{r^{4}}V_{\rm tilt}^{2} \left(z^{2}+\beta R^{2}
\right),
\label{eq:Vch_R}
\end{equation}
where 
\begin{equation}
\beta(r) =\frac{d\ln V_{\rm tilt}}
{d\ln r}.
\end{equation}

From Eq. (\ref{eq:Vch_R}), we see that if $\beta \geq 0$ (a flat rotation curve corresponds to $\beta=0$),
then $\partial V_{c,sph}^{2}/\partial R\geq 0$ at any $z$, implying that the contribution of the spherical
component to $S_{F_{R}}$ is positive or zero ($S_{F_{R}}^{sph}\geq 0$).
Moreover, in order for $\partial V_{c,sph}^{2}/\partial R$ to be negative, we
require that $\beta<0$ and $z^{2}<-\beta R^{2}$. Hence, at a fixed $R$, one can always find a
height $z_{\rm turn}=\sqrt{-\beta}R$ at which $\partial V_{c,sph}^{2}/\partial R>0$ for
any $z\geq z_{\rm turn}$. 
Since $\beta\geq -0.5$ (the case $\beta=-0.5$ corresponds to the
Keplerian decay), we have that $z_{\rm turn}\leq R/\sqrt{2}$.

For the Miyamoto \& Nagai disk with mass $M_{d}$ and
scale parameters $a$ and $b$, the radial derivative of the circular velocity is 
\begin{equation}
\frac{\partial V_{c,d}^{2}}{\partial R}=\frac{2GM_{d}R}{d^{3}}
\left[1-\frac{3}{2}\frac{R^{2}}{d^{2}}\right],
\end{equation}
where $d^{2}(R,z)\equiv R^{2}+(a+\sqrt{z^{2}+b^{2}})^{2}$.

\end{document}